\documentclass[ALICE,manyauthors]{cernphprep}

\usepackage[comma,square,numbers,sort&compress]{natbib}
\usepackage{graphicx}
\usepackage{epstopdf}
\usepackage{dcolumn}
\usepackage{bm}
\usepackage{color}
\usepackage{hyperref}
\usepackage{multirow}

\usepackage{amssymb,amsmath}
\usepackage{lineno}
\usepackage[comma,square,numbers,sort&compress]{natbib}
\usepackage{rotating}

\usepackage{textcomp}

\graphicspath{{figures/}}

\newcommand{\murm}{%
  \ifmmode
    \mathchoice
        {\hbox{\normalsize\textmu}}
        {\hbox{\normalsize\textmu}}
        {\hbox{\scriptsize\textmu}}
        {\hbox{\tiny\textmu}}%
  \else
    \textmu
  \fi
}

\begin{document}

\begin{titlepage}
\PHyear{2015}
\PHnumber{173}
\PHdate{9 July}

\title{Elliptic flow of muons from heavy-flavour hadron decays 
at forward rapidity 
in Pb--Pb collisions at $\mathbf{\sqrt{{\it s}_{\rm NN}}=2.76}$~TeV}
\ShortTitle{Heavy-flavour decay muon $v_2$ in Pb--Pb collisions}

\Collaboration{ALICE Collaboration\thanks{See Appendix~\ref{app:collab} for 
the list of collaboration members}}
\ShortAuthor{ALICE Collaboration}

\begin{abstract}
The elliptic flow, $v_{2}$, of muons from heavy-flavour hadron
decays at forward 
rapidity ($2.5 < y < 4$) is measured in Pb--Pb collisions at 
$\sqrt{s_{\rm NN}}$~=~2.76 TeV with the ALICE detector at 
the LHC.
The scalar product, 
two- and four-particle $Q$ cumulants and Lee-Yang zeros 
methods are used. 
The dependence of the $v_2$ of muons from heavy-flavour hadron decays 
on the collision centrality, in the range 0--40\%,
and on transverse momentum, $p_{\rm T}$, is studied in the 
interval $3 < p_{\rm T} < 10$~GeV/$c$. A 
positive $v_2$ is observed with the scalar product and two-particle 
$Q$ cumulants in semi-central collisions 
(10--20\% and 20--40\% centrality classes)
for the $p_{\rm T}$ interval from 3 to about 5 GeV/$c$. 
The $v_2$ magnitude tends to decrease towards more central collisions 
and with increasing $p_{\rm T}$.
It becomes compatible with zero in the interval
$6<p_{\rm T}<10~{\rm GeV/}c$.
The results are compared to models describing 
the interaction of heavy quarks and 
open heavy-flavour hadrons with the high-density medium formed in high-energy 
heavy-ion collisions. The model calculations describe 
the measured $v_2$ within uncertainties. 

\vspace*{1cm}

\end{abstract}
\end{titlepage}

\setcounter{page}{2}

\section{Introduction}\label{sec:intro}
      
Experiments with ultra-relativistic heavy-ion collisions aim at investigating 
the properties of strongly-interacting matter at very high 
temperatures and energy densities. Quantum Chromodynamics (QCD) calculations 
on the lattice predict, under these conditions, the formation of a 
Quark-Gluon Plasma (QGP), where 
colour confinement vanishes and chiral symmetry is partially
restored~\cite{Karsch:2006xs,Borsanyi:2010bp,Borsanyi:2010cj,Bazavov:2011nk,Petreczky:2013qj}.
Heavy quarks (charm and beauty) are 
created in initial hard-scattering processes 
on a time scale shorter than the QGP formation time. 
Subsequently, they interact with the medium constituents via 
inelastic~\cite{Gyulassy:1990ye,Baier:1996sk} and 
elastic~\cite{Thoma:1990fm,Braaten:1991jj,Braaten:1991we} processes. 
Therefore, heavy quarks are regarded as 
effective probes of the QGP properties. 

Heavy-quark energy loss due to in-medium interactions can be studied by means 
of the nuclear modification factor $R_{\rm AA}$, 
defined as the ratio of the yield of 
heavy-flavour particles measured in nucleus--nucleus (AA) collisions to 
that observed in proton--proton (pp)
collisions scaled by the number of binary nucleon--nucleon collisions. 
The PHENIX and STAR Collaborations measured, in central 
Au--Au collisions at $\sqrt {s_{\rm NN}}$~=~200~GeV, a strong suppression 
corresponding to a $R_{\rm AA}$ of about 0.2--0.3 for heavy-flavour decay electrons at 
mid-rapidity ($y$) and
transverse momentum $p_{\rm T} >5$~GeV/$c$~\cite{Adcox:2002cg,Adler:2004ta,Adler:2005xv,Adare:2006nq,Adare:2010de,Adare:2013yxp,Abelev:2006db}.
A similar suppression was also measured by the STAR Collaboration for mid-rapidity
${\rm D}^{0}$ mesons~\cite{Adamczyk:2014uip}.
A significant suppression was also observed by the 
PHENIX Collaboration at 
forward rapidity for muons 
from heavy-flavour hadron decays in central Cu--Cu collisions at 
$\sqrt {s_{\rm NN}}$~=~200~GeV~\cite{Adare:2012px}. 
At the LHC, the ALICE Collaboration reported a similar effect 
in central Pb--Pb collisions at $\sqrt {s_{\rm NN}}$~=~2.76 TeV 
for D mesons at mid-rapidity ~\cite{ALICE:2012ab} and 
muons from heavy-flavour hadron decays at forward 
rapidity~\cite{Abelev:2012qh} in the interval
$2<p_{\rm T}<16$~GeV/$c$ and 
$4 < p_{\rm T} < 10$~GeV/$c$, respectively.
The CMS Collaboration measured a significant 
suppression of non-prompt $\rm J/\psi$ from 
beauty-hadron decays in the interval 
$6.5 <p_{\rm T} < 30$~GeV/$c$ ($3<p_{\rm T}<30$~GeV/$c$)
and $|y|<2.4$ ($1.6<|y|<2.4$)~\cite{CMS-PAS-HIN-12-014,Chatrchyan:2012np}.
A first measurement of non-prompt ${\rm J}/\psi$ by
the ALICE Collaboration at mid-rapidity ($|y|<0.8$) and in the
interval $4.5<p_{\rm T}<10~{\rm GeV/}c$ has been recently published~\cite{Adam:2015rba}.

Further insights into 
the QGP evolution and the in-medium interactions can be gained from the 
study of the azimuthal 
anisotropy of particles carrying heavy 
quarks which, in contrast to light quarks, have experienced the full system 
evolution. 
The study of azimuthal anisotropy is a field of intense experimental 
and theoretical investigations
(see~\cite{Voloshin:2010ra} and references therein). 
In non-central collisions, 
the initial spatial anisotropy of the 
overlap region, elongated in the direction perpendicular to 
the reaction plane, defined by the beam axis and the impact parameter 
of the collision,
is converted into an anisotropy in momentum space
through rescatterings~\cite{Ollitrault:1992bk}. 
Experimentally, the study of the particle
azimuthal anisotropy is based on a 
Fourier expansion of azimuthal distributions given by:
\begin{equation}
 {{\rm d}^2N \over {{\rm d}p_{\rm T}{\rm d}\varphi }} = {1\over 2\pi} 
{{\rm d}N \over {\rm d} p_{\rm T}} 
\bigg ( 1 + 2 \sum_{n = 1}^{\infty} v_n (p_{\rm T}) {\rm cos} 
\lbrack n ( \varphi - \Psi_n) \rbrack \bigg),
\label{eq:phi}
\end{equation}
where $\varphi$ and $p_{\rm T}$ are 
the particle 
azimuthal angle and transverse momentum, 
respectively. The Fourier 
coefficients, $v_{n}$, characterize the anisotropy of produced particles and 
$\Psi_{n}$ is the azimuthal angle of the initial-state symmetry 
plane for the $n^{\rm th}$ 
harmonic, introduced to account for the event-by-event fluctuations of the 
initial nucleon density 
profile.
The second Fourier coefficient, $v_2$, which can also be expressed 
as $v_2 = \langle \rm cos \lbrack 2 (\varphi - \Psi_2) \rbrack 
\rangle$, is named elliptic flow. 

The $v_2$ of heavy-flavour hadrons is expected to provide 
information on the collective expansion 
and degree of thermalization of heavy quarks in the medium at 
low $p_{\rm T}$ ($p_{\rm T} < 2-3$~GeV/$c$). 
The participation of heavy quarks in the collective expansion
is expected to give a positive
$v_2$~\cite{Ollitrault:1992bk}. 
Moving towards intermediate $p_{\rm T}$ 
($3 < p_{\rm T} < 6$~GeV/$c$), 
the $v_2$ Fourier coefficient is also 
expected to be sensitive to the presence of recombination processes in the hadronization of 
heavy quarks~\cite{Greco:2003vf,Andronic:2003zv}. At high $p_{\rm T}$ 
($p_{\rm T} > 6$~GeV/$c$), 
the $v_2$ measurement can constrain the path-length dependence of 
the in-medium parton energy loss, which becomes the dominant 
contribution to the 
azimuthal anisotropy and is also predicted to give a 
positive $v_2$~\cite{Gyulassy:2000gk,Shuryak:2001me}, 
thus complementing the $R_{\rm AA}$ measurement.

The PHENIX Collaboration reported a positive $v_2$ of
heavy-flavour decay electrons at mid-rapidity in Au--Au collisions at 
$\sqrt {s_{\rm NN}}$~=~200~GeV, reaching a maximum value of about 0.15 at 
$p_{\rm T}$~=~1.5 GeV/$c$ in 
semi-central collisions~\cite{Adare:2006nq,Adare:2010de,Adler:2005ab}. 
A similar 
behaviour was also 
observed by the STAR Collaboration~\cite{Adamczyk:2014yew}.
Recently, a $v_2$ value significantly larger 
than zero was measured for 
D mesons at mid-rapidity in Pb--Pb collisions at 
$\sqrt {s_{\rm NN}}$~=~2.76 TeV~\cite{Abelev:2013lca,Abelev:2014ipa}. 
A complementary measurement at the same energy, 
provided by the 
heavy-flavour decay muon elliptic flow at forward 
rapidity ($2.5 < y < 4$), 
is of great interest in order to provide new 
constraints for models that implement 
the heavy-quark interactions with the medium. 
Finally, the measurement is also important for the interpretation 
of the J/$\psi$ elliptic flow results at forward rapidity~\cite{ALICE:2013xna} 
in terms of a regeneration production 
from deconfined charm quarks in the medium. 

In this Letter, we present the measurement of the elliptic 
flow of 
muons from heavy-flavour hadron decays at forward rapidity ($2.5 < y <4$)
in Pb--Pb collisions at $\sqrt {s_{\rm NN}}$~=~2.76 TeV recorded 
with the ALICE detector. 
The elliptic flow is measured using different methods: 
scalar product~\cite{Adler:2002pu}, two- and four-particle $Q$ 
cumulants~\cite{Borghini:2001vi,Bilandzic:2010jr} 
and Lee-Yang zeros~\cite{Bhalerao:2003xf,Bhalerao:2003yq,Borghini:2004ke}. 
These methods exhibit different 
sensitivities to flow fluctuations and correlations 
not related to the 
azimuthal asymmetry in the initial geometry (non-flow effects). 
The $v_2$ coefficient is measured as a 
function of $p_{\rm T}$ in the interval $3 < p_{\rm T} < 10$~GeV/$c$ 
and in three centrality 
classes in the range 0--40\%. The centrality dependence of $v_2$ 
is presented in the interval $3 < p_{\rm T} < 10$~GeV/$c$.

The Letter is organized as follows. The ALICE detector,
with an emphasis on the muon spectrometer, and the data sample are presented 
in Section~\ref{sec:ALICE}. 
The analysis details, the methods 
for the $v_2$ measurement, the inclusive muon $v_2$ determination, the 
procedure for the subtraction of the background of muons from decays of 
light-flavour hadrons and the study of systematic uncertainties, 
are described in Section~\ref{sec:strat}. 
The $v_2$ results for muons from heavy-flavour decays are presented in 
Section~\ref{sec:results} and compared to model calculations in 
Section~\ref{sec:models}.
Finally, conclusions are given in Section~\ref{sec:conc}.

\section{ALICE experiment and data sample}\label{sec:ALICE}

The ALICE detector is described in detail 
in~\cite{Aamodt:2008zz,Abelev:2014ffa}. 
The apparatus is 
composed of a set of central barrel detectors (pseudo-rapidity coverage 
$\vert \eta \vert < 0.9$) located inside a solenoid magnet that 
generates a field of 0.5~T parallel to the beam direction, 
a muon spectrometer ($ -4 < \eta < -2.5$\footnote{In the ALICE reference 
frame, the muon spectrometer covers a negative $\eta$ range and consequently a 
negative $y$ range. In the following, given that the colliding system is 
symmetric, the results are presented with a 
positive $y$ notation.}) and a set of detectors for event 
characterization and triggering located in the forward and 
backward $\eta$ regions. 
The muon spectrometer consists of a passive front 
absorber made of carbon, concrete and steel, a beam shield, 
a 3~T$\cdot$m dipole magnet, 
tracking chambers, a muon 
filter (iron wall) and trigger chambers. The muon tracking system is 
composed of five stations, each including two 
planes of cathod pad chambers, 
with the third station 
inside the dipole magnet. The muon tracking system is completed by four 
trigger planes of resistive plate chambers downstream of the iron wall, 
which absorbs hadrons that punch through the front absorber, 
as well as secondary particles produced inside it 
and low momentum muons ($p < 4$~GeV/$c$). 
Two scintillator arrays (V0) covering the 
pseudo-rapidity intervals $-3.7 < \eta < -1.7$ and $2.8 < \eta < 5.1$ are used 
for triggering, for collision centrality determination and 
for beam-induced background rejection. The Zero 
Degree Calorimeters (ZDC), located at 114~m from the centre of the 
detector on both sides,
can detect spectator 
protons and neutrons and are also 
used for the offline rejection of beam-induced background and 
electromagnetic interactions. 
The Silicon Pixel Detector (SPD), that 
composes the two innermost layers of the Inner Tracking System (ITS), 
is used 
for the interaction vertex reconstruction.
The Time Projection Chamber (TPC), which measures charged-particle tracks 
with full azimuthal coverage in $\vert \eta \vert < 0.9$, is used in 
this analysis for the measurement of the reference particles (Section 3.1).

The results presented in this Letter are obtained from
the data sample recorded 
with ALICE during the 2011 Pb--Pb run. 
The data were collected with a minimum-bias 
trigger requiring the coincidence of signals in the two V0 arrays in 
synchronization with the passage of two crossing bunches. 
In addition, the 
recorded event sample was enriched with central and 
semi-central Pb--Pb collisions by applying thresholds, 
at the trigger level, 
on the V0 signal amplitude. The 
beam-induced background (beam-gas interactions) was reduced by using the 
timing information from the 
V0 and ZDC detectors. Furthermore, a minimal energy deposit in the ZDC 
was required to reject the contribution from electromagnetic Pb--Pb 
interactions. 
Only events with a reconstructed primary vertex within $\pm$~10~cm 
from the nominal position of the interaction vertex along the beam direction 
are analyzed.
The Pb--Pb collisions are classified according to 
their degree of centrality 
by means of the sum of the amplitudes of the signals in the V0 
detector and the centrality classes are defined as 
percentiles of the total 
hadronic Pb--Pb cross section~\cite{Abelev:2013vea}. 
The analysis is carried out in three centrality classes: 
0--10\%, 10--20\% and 20--40\%. The analysed data sample
corresponds to an integrated 
luminosity of 11.3~$\murm\rm{b}^{-1}$ in the 0--10\% centrality class and of 
3.5~$\murm\rm{b}^{-1}$ in the other two centrality classes.

\section{Data analysis}\label{sec:strat}

The elliptic flow of muons from heavy-flavour hadron decays, 
$v_2^{\mu \leftarrow {\rm HF}}$, is obtained from the measurement of the 
inclusive muon elliptic flow, $v_2^{\mu}$, by subtracting the elliptic flow 
of muons from primary charged pion and kaon decays 
$v_2^{\mu  \leftarrow \pi, {\rm K}}$
(Sections~\ref{sec:trackcut} and~\ref{sec:MuPiK}), as:
\begin{equation}
v_2^{\mu \leftarrow {\rm HF}}  = 
{v_2^{\mu} - f^{\mu \leftarrow \pi, {\rm K}} \cdot 
v_2^{\mu \leftarrow \pi, {\rm K}} \over 
{ 1 - f^{\mu \leftarrow \pi, {\rm K}}  }},
\label{eq:v2HFdef}
\end{equation}
where $f^{\mu \leftarrow \pi,K}$ is the muon background fraction, 
defined as the ratio of the yield of muons from primary charged pion and kaon 
decays to that of inclusive muons. 
The measurement of the $v_2^{\mu \leftarrow {\rm HF}}$ coefficient is 
carried out in the interval $3 <p_{\rm T} < 10$~GeV/$c$ 
in order to limit the systematic uncertainty on the subtraction of the 
muon background contribution.

\subsection{Track selection}\label{sec:trackcut}

The selection criteria for particles of interest, muon tracks, 
are similar to those used in the previous analyses of pp collisions at 
$\sqrt s$~=~2.76~TeV and 7 TeV and Pb--Pb collisions at 
$\sqrt {s_{\rm NN}}$~=~2.76~TeV~\cite{Abelev:2012qh,Abelev:2012pi}. 
The tracks 
are required to be within 
the geometrical acceptance of the muon spectrometer, with 
$-4 < \eta < -2.5$ and $170^\circ < \theta_{\rm abs} < 178^\circ$, 
where $\theta_{\rm abs}$ is the polar angle measured at the end of the 
absorber. In order to improve the muon identification, 
a reconstructed track in the 
tracking chambers is required to match a track segment in the trigger 
chambers. 
This leads to a very efficient rejection of the background produced 
by charged hadrons, which are absorbed in the iron wall. 
Furthermore, a cut on the product $p \cdot DCA$ of the 
track momentum $p$ and distance of closest approach ($DCA$) 
to the primary vertex 
is applied to remove the beam-induced background tracks 
and fake tracks 
coming from the superposition of several particles crossing the muon 
spectrometer. This cut is set to $6\sigma$, 
where $\sigma$ is extracted from a Gaussian fit to the 
$ p\cdot DCA$ distribution measured in two intervals of $\theta_{\rm abs}$, 
corresponding to different materials in the front absorber. 
After the cuts are applied, 
in the region $p_{\rm T} > 3$~GeV/$c$ the residual 
background to heavy-flavour decay muons consists of muons 
from decays of primary charged pions and 
kaons\footnote{Note that the contribution of muons from 
secondary light hadron 
decays produced inside the front absorber is negligible 
for $p_{\rm T}>3$~GeV/$c$~\cite{Abelev:2012pi}.}
and it amounts to 5--15\%, 
depending on $p_{\rm T}$ and on collision centrality 
(Section~\ref{sec:MuPiK}).

The mid-rapidity charged-particle tracks used to determine the flow vector 
$\vec Q_n$ or the generating function (Section~\ref{sec:flowmeth}) are 
called 
in the following reference particles. 
They are defined as tracks measured in the 
TPC in $\vert \eta \vert < 0.8$. 
These are required to have at least 70 
associated space points out of the maximum of 159, 
a $\chi^2$ per degree of freedom (ndf) for the momentum fit 
in the range $\chi^2/{\rm ndf} < 2$ and a transverse momentum value in the 
interval $0.2 < p_{\rm T} < 5$~GeV/$c$. Additionally, tracks are rejected if 
their distance of closest approach to the primary vertex is larger than 3~cm 
in the plane transverse to the beam direction or in the 
longitudinal direction.

\subsection{Flow analysis methods}\label{sec:flowmeth}

The elliptic flow measurement is carried out using various 
methods 
that have different sensitivities to flow fluctuations
and non-flow effects~\cite{ALICE:2011ab}.
Flow fluctuations
are mainly due to event-by-event 
fluctuations of the initial density profile,
while non-flow effects correspond to correlations 
not related to the 
azimuthal anisotropy in the initial state, such as 
resonance decays, jets and 
Bose-Einstein correlations between identical particles. 
It is worth mentioning that, in the present 
analysis, most of these non-flow effects 
are strongly suppressed by introducing an $\eta$ gap 
between reference particles 
and particles of interest~\cite{Luzum:2010sp}.
In this analysis, the scalar product~\cite{Adler:2002pu},
two- and four-particle 
$Q$ cumulants~\cite{Borghini:2001vi,Bilandzic:2010jr}
and Lee-Yang 
zeros~\cite{Bhalerao:2003xf,Bhalerao:2003yq,Borghini:2004ke} methods 
are employed. The description of these methods will be limited to the 
features specific 
to the present analysis. The following notations are 
introduced: 
$v_2^{\mu (\mu \leftarrow {\rm HF})} \lbrace {\rm SP} \rbrace$,
refers to the measurement using the scalar product,
$v_2^{\mu (\mu \leftarrow {\rm HF})} \lbrace {\rm 2} \rbrace$ 
and
$v_2^{\mu (\mu \leftarrow {\rm HF})} \lbrace {\rm 4} \rbrace$ 
correspond to the ones using the two-particle $Q$ cumulants 
and four-particle $Q$ cumulants, while
$v_2^{\mu (\mu \leftarrow {\rm HF})} \lbrace {\rm LYZ-Prod} \rbrace$ and 
$v_2^{\mu (\mu \leftarrow {\rm HF})} \lbrace {\rm LYZ-Sum} \rbrace$ 
are obtained using Lee-Yang zeros with product and sum generating functions.
The superscripts $\mu$ and $\mu \leftarrow {\rm HF}$ 
refer to inclusive muons and muons from heavy-flavour hadron
decays, respectively. 
It is worth mentioning that these methods are 
more accurate than the standard event plane 
method, which yields a measurement lying between the event-averaged mean value 
and the root-mean-square value in the presence of flow 
fluctuations~\cite{Poskanzer:1998yz,Luzum:2012da}. 
Moreover, the multi-particle correlation methods 
(four-particle $Q$ cumulants and Lee-Yang zeros) are less affected 
by non-flow correlations than two-particle correlation methods, 
but they cannot be used reliably when the muon flow 
magnitude is small and when the number of 
muons is small 
in the selected phase-space region e.g. in 
central and peripheral 
collisions, respectively~\cite{Borghini:2001vi,Bhalerao:2003xf}. 
Under these conditions, 
the scalar product and two-particle cumulant methods 
provide a $v_2$ measurement in a wider centrality range.

The scalar product method~\cite{Adler:2002pu,Poskanzer:1998yz}, derived 
from the standard event 
plane technique~\cite{Poskanzer:1998yz}, is based on the measurement of the 
flow vector $\vec Q_n$~\cite{Adler:2002pu} computed 
from reference particles. In order to 
determine the elliptic flow, the $\vec Q_2$ vector in a given event
is expressed as: 
\begin{equation}\label{eq:Qvector}
\vec Q_2 = \Big(\sum_{j = 1}^{N} {\rm cos} 2\varphi_j, 
\sum_{j = 1}^{N} {\rm sin} 2\varphi_j\Big), 
\end{equation}
where $\varphi_j$ is 
the particle azimuthal angle and $N$ is the multiplicity of 
reference particles.
 
With this method the $2^{\rm nd}$ harmonic coefficient is given by:
\begin{equation}
v_2\lbrace {\rm SP} \rbrace = 
{\langle \vec Q_2 \cdot \vec u_{2, i} (\eta, p_{\rm T}) \rangle \over 
{2 \sqrt {\langle \vec{Q_2^{\rm A}} \cdot \vec{Q_2^{\rm B}}\rangle}}}, 
\label{eq:SP}
\end{equation}
where the brackets in the numerator indicate the average over muons 
at forward rapidity, in all events. 
The vector $Q_2$ is calculated from Eq.~(\ref{eq:Qvector}) and the vector 
$\vec u_{2, i} = 
({\rm cos} 2 \varphi_{i}, {\rm sin} 2 \varphi_{i})$
is the unit vector of the 
$i^{\rm th}$ muon.  
In the denominator, each sample of reference particles
used to compute $\vec Q_2$ is divided 
into two sub-samples of same multiplicity 
in symmetrical $\eta$ intervals, $-0.8 < \eta < -0.5$ and 
$0.5 < \eta <0.8$, 
separated by a $\eta$ gap of one unit of pseudo-rapidity, labeled with the 
superscripts $\rm A$ and ${\rm B}$ and the brackets correspond to 
the average over events.

The cumulant technique~\cite{Borghini:2001vi,Bilandzic:2010jr} is 
based on a cumulant expansion of multi-particle azimuthal correlations. 
Different order cumulants have different sensitivities to flow fluctuations. 
In the present analysis, two- and four-particle cumulants are used 
to extract the muon elliptic flow. 
The results presented in the following are obtained from 
a direct calculation of multi-particle cumulants performed by
using the $Q$-cumulant 
technique~\cite{Bilandzic:2010jr}, which is based on the moments of the 
magnitude of the flow vector $\vec{Q_2}$.
It is worth mentioning that in this approach the 
cumulants are not biased by the interferences between various harmonics. 
The reference elliptic flow values $V_2$ evaluated from 
the 2$^{\rm nd}$ order cumulant $c_2\lbrace 2 \rbrace$ and 
$4^{\rm th}$ order cumulant $c_2\lbrace 4 \rbrace$ with reference particles 
are given by 
$V_ 2\lbrace 2 \rbrace = \sqrt {c_2\lbrace 2 \rbrace}$ and 
$V_ 2\lbrace 4 \rbrace =\ ^4\sqrt {- c_2\lbrace 4 \rbrace}$, respectively. 
Once the reference elliptic flow is estimated, the muon 
elliptic flow with respect to the reference elliptic flow is 
obtained from the
2$^{\rm nd}$ and 4$^{\rm th}$ order cumulants according to:
\begin{equation}\label{eq:diffQC2}
v_2 \lbrace 2 \rbrace  = \frac {d_2 \lbrace 2 \rbrace} 
{V_ 2\lbrace 2 \rbrace} 
\ {\rm and}\ 
v_2 \lbrace 4 \rbrace  = \frac {d_2 \lbrace 4 \rbrace} 
{V_ 2\lbrace 4 \rbrace^3}, 
\end{equation}

where $d_2\lbrace 2 \rbrace$ and $d_2\lbrace 4 \rbrace$ are the 
$2^{\rm nd}$ and $4^{\rm th}$ order cumulants of 
selected muons~\cite{Bilandzic:2010jr}.

The Lee-Yang zeros 
method~\cite{Bhalerao:2003xf,Bhalerao:2003yq,Borghini:2004ke}
relies on 
correlations involving all particles in the event. This is the 
limit of cumulants when the order of cumulants
goes to infinity. The method is 
based on the location of the zeros in the complex plane, of a 
generating 
function of azimuthal correlations, which relates the position of 
the first minimum of the generating function to the magnitude of the 
reference elliptic flow $V_2$ defined as:
\begin{equation}\label{eq:refv2}
V_2 \equiv \bigg \langle \sum_{j = 1}^M {\rm cos} [2 (\varphi_j - \Psi_2)] 
\bigg\rangle_{\rm events},
\end{equation}
where $M$ is the multiplicity of reference particles and the average is 
taken over all events. For this purpose, the following 
complex-valued generating 
function is evaluated as a function of a positive real variable $r$ 
and few, typically five,
equally spaced reference angles $\vartheta$
($\rm {LYZ - Prod}$ method):
\begin{equation}\label{eq:gene1}
G^\vartheta (ir) \equiv \bigg \langle \prod_{j = 1}^M ( 1 + ir {\rm cos} 
[2(\varphi_j - \vartheta)]) \bigg \rangle_{\rm events}.
\end{equation}
The first positive minimum of $\vert G^\vartheta(ir) \vert$,
denoted as $r_0^\vartheta$,
allows one to estimate $V^\vartheta_2$,
which can be written as 
$V_2^\vartheta = j_{01}/r_0^\vartheta$, 
where $j_{01} \simeq 2.405$ is the first root 
of the Bessel function. Once the first minimum $r^\vartheta_0$ is determined, 
the differential muon elliptic flow is estimated with respect to the reference 
flow $V_2^\vartheta$ as detailed in~\cite{Borghini:2004ke}. 
Finally, the result is averaged over 
all $\vartheta$ angles. An alternative form of the generating
function
provided with the $\rm{LYZ-Sum}$ method is:
\begin{equation}\label{eq:gene2}
G^\vartheta (ir) \equiv \bigg \langle {\rm exp} \bigg (ir \sum_{j = 1}^M 
{\rm cos} [ 2 (\varphi_j - \vartheta)] \bigg )\bigg \rangle_{\rm events}. 
\end{equation}
The version of the method involving a product for the construction of the 
generation function (Eq.~(\ref{eq:gene1})) 
was designed to disentangle 
interferences between different harmonics, which is not the case with 
the generating function using a sum of the individual reference particle 
contributions. Both generating functions are used in this analysis. 

Note that, for all methods, autocorrelation effects are avoided 
because the particles (muons) used in the determination 
of the flow are not included 
in the estimation of the reference flow. 

\subsection{Inclusive muon elliptic flow}\label{sec:inclv2}

The elliptic flow of inclusive muons, $v_2^\mu$, 
is studied with two-particle 
correlation methods (scalar product and two-particle $Q$ cumulants) in the centrality intervals 0--10\%, 10--20\%
and 20--40\%. In the 20--40\% centrality interval, 
the multi-particle correlation methods (four-particle $Q$ cumulants and Lee-Yang zeros)
are also used. 

\begin{figure}[!th]
\centering
\includegraphics[width=.49\textwidth]{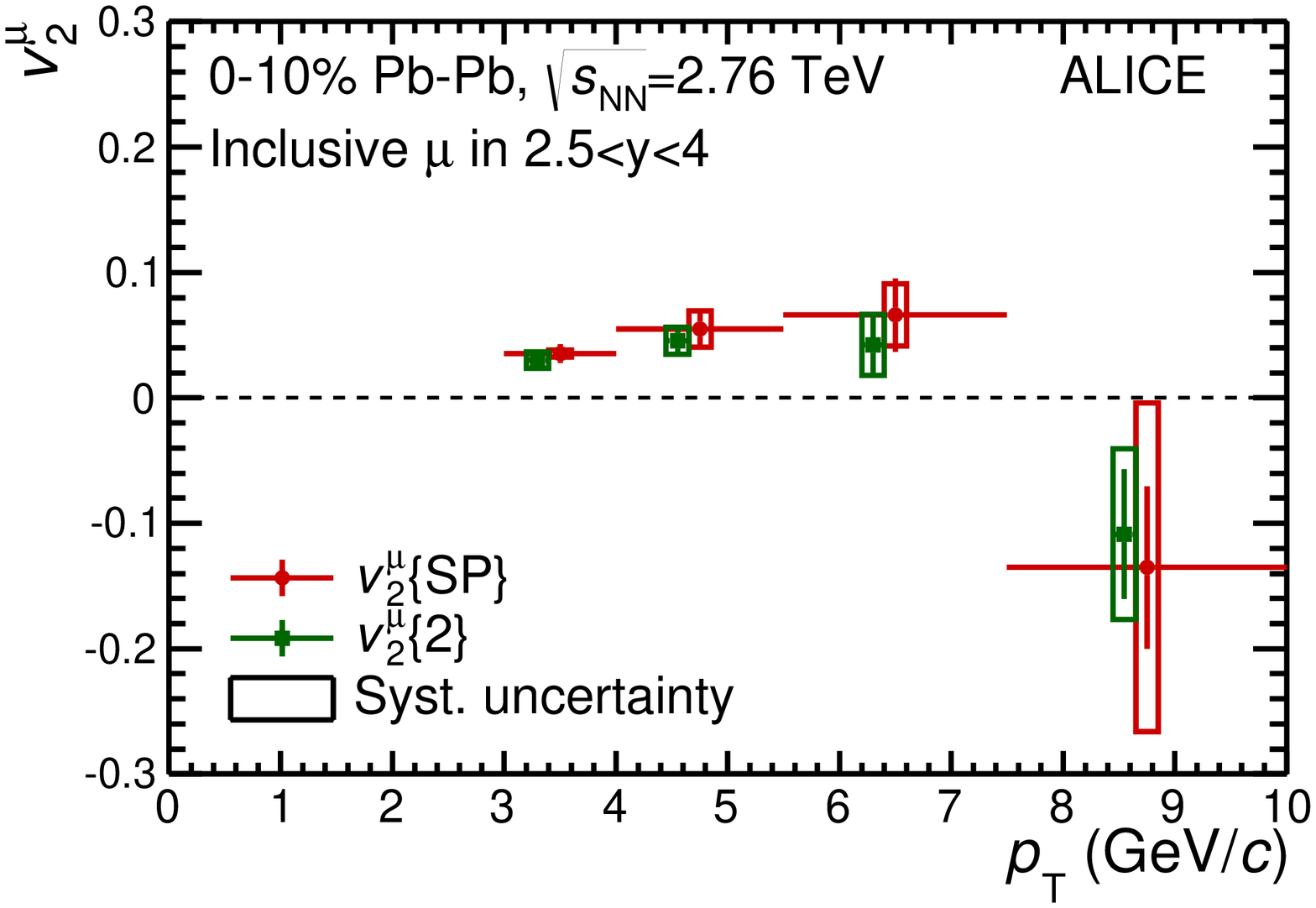}
\includegraphics[width=.49\textwidth]{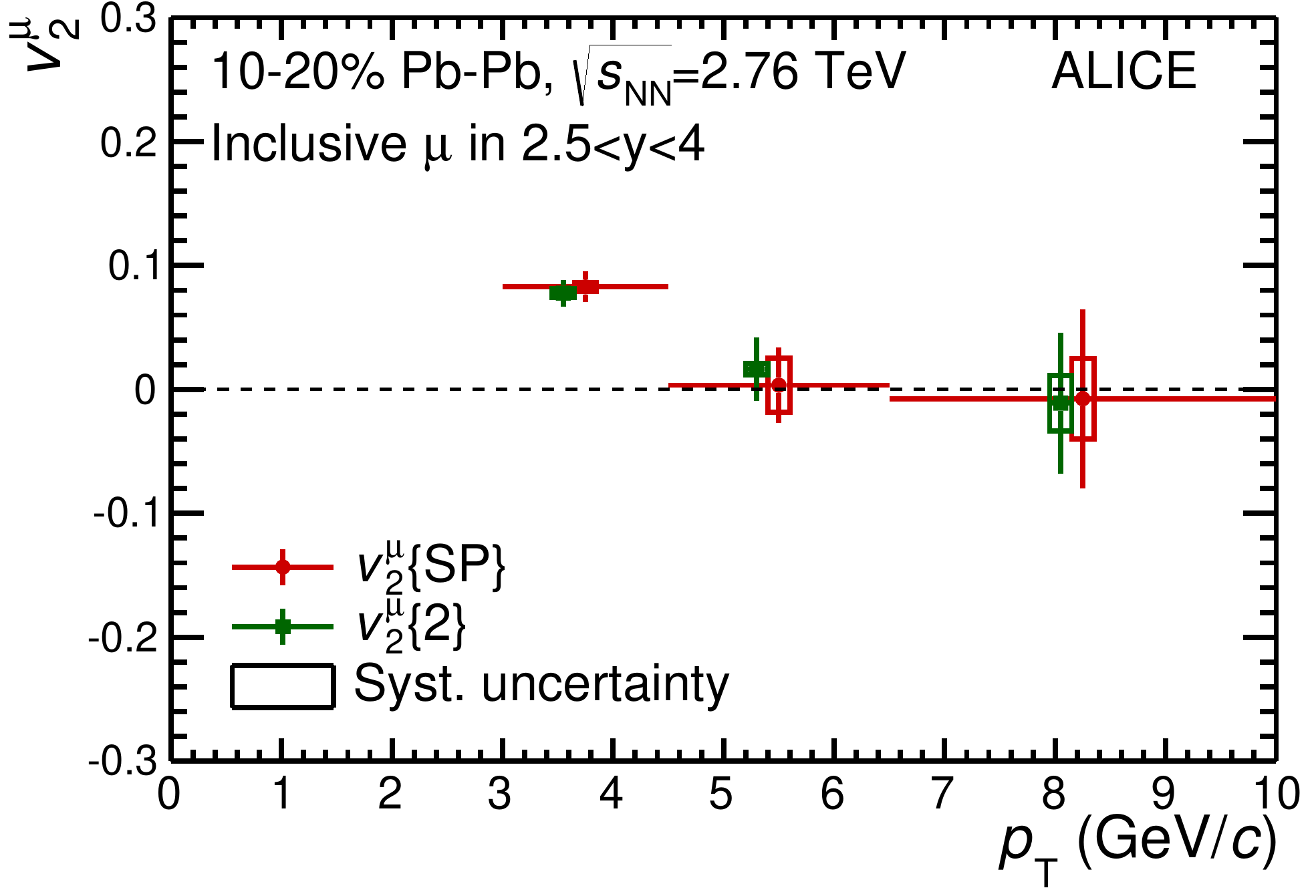}
\includegraphics[width=.49\textwidth]{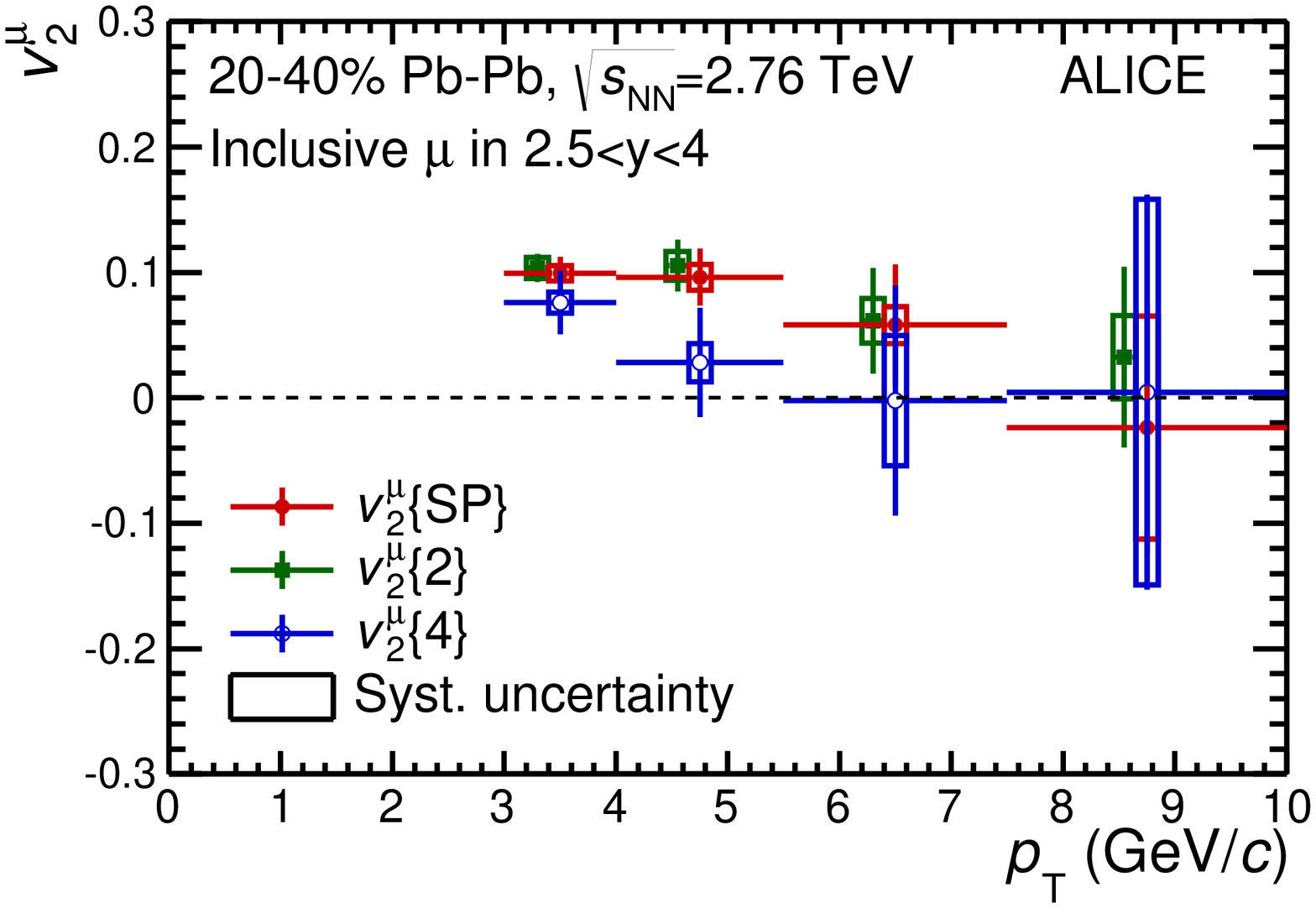}
\includegraphics[width=.49\textwidth]{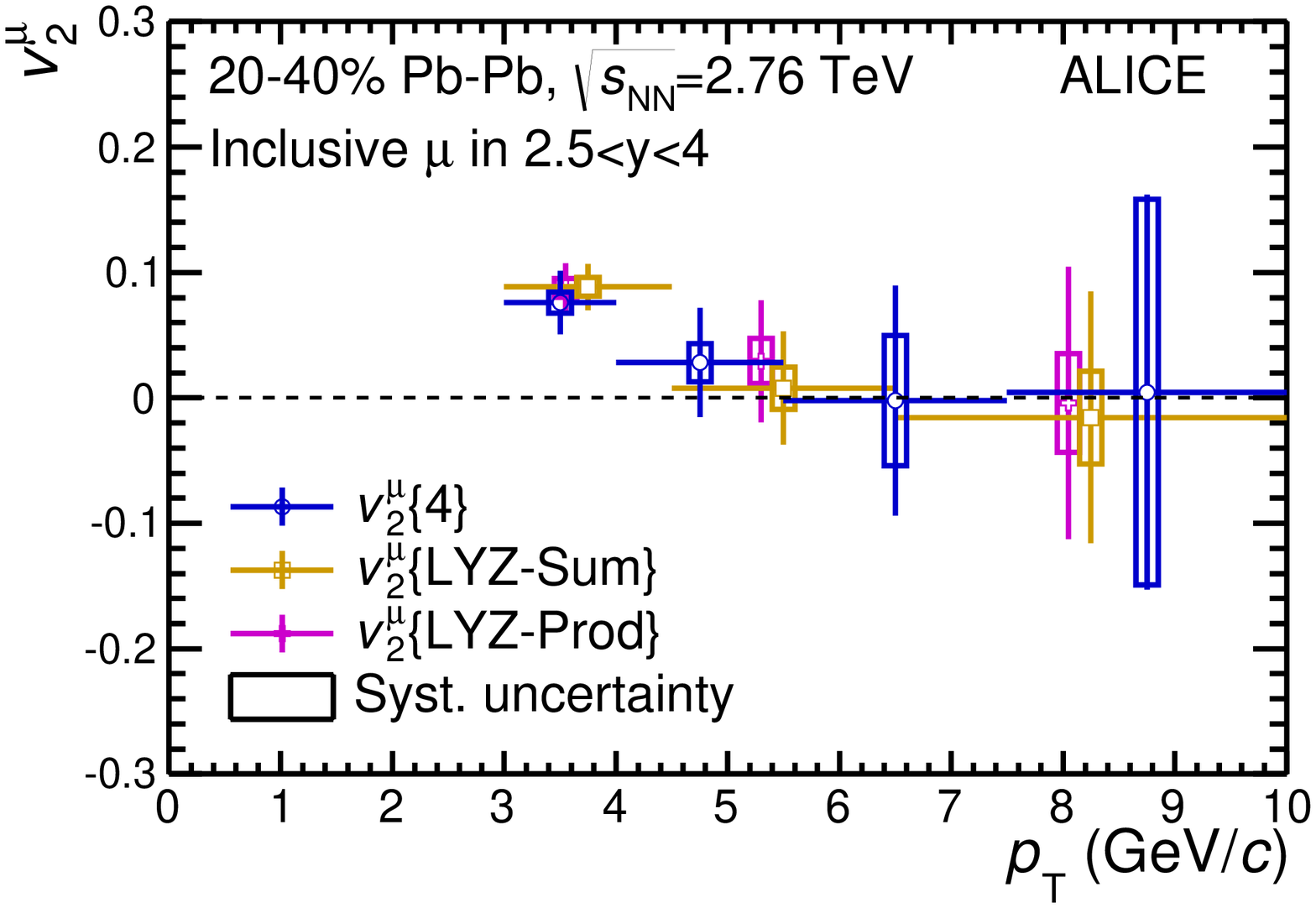}
\caption{$p_{\rm T}$-differential inclusive muon $v_2$ in $2.5 < y <4$ 
and various centrality intervals, in Pb--Pb collisions at 
$\sqrt {s_{\rm NN}}$ = 2.76 TeV. The symbols are placed at the centre of 
the $p_{\rm T}$ interval and, for visibility, the points from 
two-particle $Q$ 
cumulants and Lee-Yang zeros with product generating function 
are shifted horizontally. The vertical error bars represent the statistical 
uncertainty, the horizontal error bars correspond to the 
width of the bin (not shown for the shifted data points) and the open boxes are the systematic 
uncertainties.
The $p_{\rm T}$ intervals used with the Lee-Yang zeros 
method are different with respect to the other methods.
Upper panels: results from two-particle 
correlation flow methods (scalar product and two-particle $Q$ cumulants) 
in the 0--10\% (left) and 10--20\% (right) centrality intervals. 
Lower panels: results in the 20--40\% centrality 
interval from two-particle 
correlation flow methods (scalar product and two-particle $Q$ cumulants) and 
from four-particle $Q$ cumulants (left), and from 
four-particle $Q$ cumulants and Lee-Yang zeros (right).}
\label{fig:v2Mu}
\end{figure}

Several sources of systematic uncertainty affecting the muon elliptic flow 
measurement are considered. 
These take into account the changes due to the 
variations of the reference particle selection criteria as 
in~\cite{Abelev:2013lca,Abelev:2014ipa,Abelev:2012di}, 
to allow us to check the 
robustness of the $v_2^\mu$ measurement. 
Since the collision impact parameter distribution could slightly 
depend on the observable used for the 
centrality determination, a systematic uncertainty is
estimated by 
repeating the analysis using the number of clusters in 
the outermost layer of the SPD and the number of tracks 
in the TPC as 
centrality estimators, instead of the V0 signal amplitude.
The systematic uncertainty due to the effect of 
TPC tracks from different Pb--Pb collisions piled-up in the 
same recorded event is estimated 
by applying a tighter cut to remove outliers in the multiplicity 
distribution of reference particles. 
This is done by requiring that the centrality 
values determined
using the V0 signal amplitude and the number of TPC tracks 
do not differ by more than 5\%.
An additional systematic uncertainty specific to the scalar product 
is evaluated by varying the $\eta$ gap 
between the two sub-events from 1 to 0.8 $\eta$-units
(see Eq.~(\ref{eq:SP}) and~\cite{Adler:2002pu}). 
The various systematic uncertainties are added in quadrature. 
They tend to increase with increasing $p_{\rm T}$ (see 
Fig.~\ref{fig:v2Mu}).
A summary of the systematic 
uncertainties, in the interval $3 < p_{\rm T} < 4.5$~GeV/$c$, is 
presented in Table~1. 

\begin{table}[hbt!]\label{tab:systv2mu}
\centering
\begin{tabular}{ c|c|c|c|c }\hline
\multirow{2}{*}{$v_2^\mu$ analysis} & \multirow{2}{*}{Source}
& \multicolumn{3}{c}{Systematic uncertainty ($\%$)} \\
\cline{3-5}
& & 0--10\%  & 10--20\%  & 20--40\%  \\ \hline
$v_2^\mu \lbrace {\rm SP} 
\rbrace$ & Reference particles & 3 & 1 & 3   \\
 & Centrality selection & 6 & 1 &4\\
 & TPC pile-up & 2& 4 & 2 \\
 & $\eta$ gap & 13 & 1  & 1 \\ \hline
 $v_2^\mu \lbrace 2 \rbrace$ 
&  Reference particles & 13 & 3  & 2  \\ 
 & Centrality selection  & 14 & 3 & 6   \\
 & TPC pile-up & 8  & 1  & 4  \\ \hline
$v_2^\mu \lbrace 4 \rbrace$ 
& Reference particles & & & 10  \\ 
 & Centrality selection & & & 1  \\
 & TPC pile-up & & & 1\\  \hline
$v_2^\mu \lbrace {\rm LYZ-Sum} \rbrace$ 
& Reference particles & & & 4 \\ 
 & Centrality selection &  & & 7 \\
 & TPC pile-up & & & 2  \\  \hline
$v_2^\mu \lbrace {\rm LYZ-Prod} \rbrace$ 
&Reference particles & & & 2 \\ 
 & Centrality selection  & & & 8  \\
 & TPC pile-up & & & 2  \\  \hline
\end{tabular}
\caption{Systematic uncertainty sources affecting the 
inclusive muon elliptic flow measurement in the 0--10\%, 10--20\% and 
20--40\% centrality classes for the interval $3 < p_{\rm T} < 4.5$~GeV/$c$. 
They are given as a percentage of the $v_2$ value. }
\end{table}

Figure~\ref{fig:v2Mu} shows the $p_{\rm T}$-differential muon elliptic flow 
($v_2^\mu$) in the 0--10\%, 10--20\% and 20--40\% centrality classes 
as obtained using the various methods.
The values of $v_2^\mu$ slightly increase from central to semi-central 
collisions and this effect is 
more pronounced in the $p_{\rm T}$ interval
$2 < p_{\rm T} < 4$~GeV/$c$. 
The two-particle correlation methods (scalar product and two-particle $Q$ 
cumulants) give consistent 
results over the whole $p_{\rm T}$ range, indicating  
that these methods have a similar sensitivity to 
non-flow effects\footnote
{Note that, in this analysis, most 
non-flow correlations are suppressed, even with two-particle 
correlation methods since reference particles and 
inclusive muons are separated by at 
least 1.7 $\eta$-units. However, it is worth mentioning that the main 
difference between the two methods is the $\eta$ gap between the two 
sub-samples used to compute $\vec{Q}_2$ (Eq.~(\ref{eq:SP})) which also allows 
to partly remove non-flow effects.} and 
in particular to flow fluctuations. A similar 
agreement is found when comparing the multi-particle correlation methods 
(four-particle $Q$ cumulants and Lee-Yang zeros) to each other. 
No significant difference between the $v_2^\mu$ results 
extracted with Lee-Yang zeros using either the sum or product generating 
function is seen, hence indicating that 
interferences between harmonics are negligible~\cite{Abelev:2008ae}.
The four-particle $Q$ cumulants and Lee-Yang zeros are expected to be less affected by non-flow effects than scalar product 
or two-particle 
$Q$ cumulants~\cite{Chatrchyan:2012ta}. Moreover, since four-particle $Q$ 
cumulants 
give comparable results as Lee-Yang zeros, one can conclude 
that non-flow correlations are almost completely removed 
at 
the $4^{\rm th}$ order. Finally, the central values of $v_2^\mu$ obtained with 
four-particle $Q$ cumulants or 
Lee-Yang zeros are systematically smaller than with two-particle 
correlation methods, although compatible within uncertainties.  
Such differences may indicate that initial 
fluctuations play a role in 
the development of the final momentum-space anisotropy. 

\subsection{Muon background subtraction}\label{sec:MuPiK}

The subtraction of the muon background contribution to the measured $v_2^\mu$ 
requires an estimate of the elliptic flow of muons from 
charged pion and kaon decays, $v_2^{\mu \leftarrow {\rm \pi,K}}$, and of the 
background fraction, $f^{\mu \leftarrow {\rm \pi,K}}$ 
(see Eq.~(\ref{eq:v2HFdef})).
The determination of the $v_2^{\mu \leftarrow {\rm \pi,K}}$ 
coefficient requires two steps.
First, the $p_{\rm T}$- and $\eta$-differential 
$v_2$ of charged particles measured in $\vert \eta \vert < 2.5$ by the ATLAS 
Collaboration in Pb--Pb collisions~\cite{ATLAS:2011ah}
and the $p_{\rm T}$ distributions of 
charged pions and kaons measured in $\vert y \vert < 0.8$ by the 
ALICE Collaboration in pp and Pb--Pb collisions~\cite{Christiansen::pc2013,Abelev:2014laa} are extrapolated to forward rapidity.
Then, the $p_{\rm T}$ distributions of muons from charged pion and kaon decays, 
needed to estimate $f^{\mu \leftarrow {\rm \pi,K}}$ and 
$v_2^{\mu \leftarrow {\rm \pi,K}}$, are generated according to a 
simulation taking into account 
the decay kinematics and the effect of the front absorber.

The $p_{\rm T}$- and $\eta$-differential elliptic flow of charged particles in 
$\vert \eta \vert < 2.5$, $v_2^{\rm ch}$, is extrapolated to forward 
rapidity using:
\begin{equation}
v_2^{\rm ch}(p_{\rm T}, \eta) = F(\eta) \cdot 
v_2^{\rm ch}(p_{\rm T}, 2 < \vert \eta \vert < 2.5),
\label{eq:v2etrap}
\end{equation}
where $v_2^{\rm ch}(p_{\rm T}, 2 < \vert \eta \vert < 2.5)$ is the measured 
charged-particle elliptic flow 
in $2 < \vert \eta \vert < 2.5$ with 
the event plane method. 
Since the $v_2^{\rm ch}(p_{\rm T})$ measured by the 
ATLAS Collaboration is affected by statistical 
fluctuations, it is assumed that in the interval 
$10 < p_{\rm T} < 20$ GeV/$c$, needed to simulate the decay muons 
up to $p_{\rm T}$ = 10~GeV/$c$, $v_2^{\rm ch}$ remains constant with a value 
given by the one measured in the interval $10 < p_{\rm T} <12$~GeV/$c$. 
The extrapolation factor $F(\eta)$ is calculated by parameterizing 
the $\eta$-differential $v_2^{\rm ch}$ measured by the ATLAS 
Collaboration in 
various $p_{\rm T}$ intervals with a second order polynomial. In 
the interval $7<p_{\rm T}<20$~GeV/$c$, 
the ATLAS $v_2^{\rm ch}$
does not show a dependence on $\eta$ 
in $\vert \eta \vert < 2.5$. 
Therefore, for $p_{\rm T} > 7$~GeV/$c$,
$F(\eta)$ is computed as the average between 
a flat extrapolation 
function and the extrapolation factor obtained with the parabolic 
parameterization in $4 <p_{\rm T} < 7$~GeV/$c$.

The mid-rapidity charged pion and kaon $p_{\rm T}$
distributions measured in Pb--Pb 
collisions are extrapolated to forward rapidity using the same strategy as 
in~\cite{Abelev:2012qh} and summarized in the following. 
Assuming that the 
nuclear modification factor $R_{\rm AA}^{\pi, \rm K}$ of charged pions 
and kaons in Pb--Pb collisions does not depend on 
rapidity up to $y$~=~4~\cite{Abelev:2012qh,Aad:2015wga},
the $p_{\rm T}$ distributions of 
charged pions and kaons at forward rapidity can be expressed as: 
\begin{equation}  \
\frac{{\rm d}N_{\rm {PbPb}}^{\pi, {\rm K}}} 
{{\rm d}p_{\rm T}{\rm d} y } =  
\langle T_{\rm AA} \rangle  \cdot 
\frac {{\rm d}\sigma_{\rm {pp}}^{\pi,{\rm K}} }
{{\rm d}p_{\rm T}{\rm d} y } \cdot
[R_{\rm AA}^{\pi,{\rm K}}(p_{\rm T})]_{y = 0},
\label{eq:HadronPbPbFW}
\end{equation}
where $\langle T_{\rm AA} \rangle$ is the average nuclear overlap
function in centrality classes under study, estimated as described in~\cite{Abelev:2013qoq}.
The systematic uncertainty introduced by the assumption on 
$R_{\rm AA}^{\pi, \rm K}$ will be discussed later.
The rapidity extrapolation of the mid-rapidity pion and kaon 
$p_{\rm T}$-differential cross sections
measured in pp 
collisions~\cite{Abelev:2012qh,Adler:2006yu} is done according to:
\begin{equation}\label{eq:extrapol}  
\frac{{\rm d^2}\sigma_{\rm {pp}}^{\pi, {\rm K}} }
{{\rm d}p_{\rm T} {\rm d}y} = 
 \bigg\lbrack \frac {{\rm d^2}\sigma_{\rm {pp}}^{\pi, {\rm K}} }
{{\rm d}p_{\rm T} {\rm d} y} \bigg\rbrack_{y = 0}               
\cdot {\rm exp} \Big({-y^2 \over {2 \sigma^2_y}}\Big),                  
\end{equation}
$\sigma_y$ being estimated from Monte-Carlo event generators 
(see~\cite{Abelev:2012qh} for details).

The elliptic flow of muons from charged 
pion and kaon decays, $v_2^{\mu \leftarrow \pi, {\rm K}}$, in 
$2.5 < y < 4$ and in various centrality 
classes\footnote{The $v_{2}^{\mu\leftarrow\pi,{\rm K}}$ of muons 
from charged pion and kaon
decays in the 20--40\% centrality class is then obtained from the mean 
of the charged-particle $v_{2}$ in 20--30\% 
and 30--40\% centrality classes,
with an additional systematic uncertainty
provided by the difference with respect to the results in these two 
centrality classes.}, is obtained by 
means of fast 
simulations using $v_2^{\rm ch}(\eta, p_{\rm T})$ given by 
Eq.~(\ref{eq:v2etrap}) 
and charged pion and kaon $p_{\rm T}$ distributions 
as obtained from Eq.~(\ref{eq:HadronPbPbFW})--(\ref{eq:extrapol}). 
The absorber effect is accounted for by rejecting the pions and kaons that 
do not decay within a distance corresponding to one interaction length 
from the beginning of the absorber. 
The simulation was repeated twice, considering that 
charged particles are either all pions or all kaons. 
The estimated $v_2^{\mu \leftarrow \pi, {\rm K}}$ 
decreases 
with increasing $p_{\rm T}$ from about 0.085 at $p_{\rm T}$~=~3~GeV/$c$ 
to 0.035 at $p_{\rm T}$~=~10~GeV/$c$ in 
the 10\% most central 
collisions. In the 20--40\% centrality interval, 
the $v_2^{\mu \leftarrow \pi, {\rm K}}$ values vary between 
0.19 ($p_{\rm T}$~=~3~GeV/$c$) and 0.08 
($p_{\rm T}$~=~10~GeV/$c$)

The background fraction, $f^{\mu \leftarrow {\rm \pi, K}}$, is calculated 
as the ratio of the $p_{\rm T}$-differential yield of muons from charged 
pion and kaon decays in $2.5 < y < 4$ obtained in the simulation 
to the 
measured $p_{\rm T}$-differential yield of inclusive muons. 
It decreases as $p_{\rm T}$ increases, 
from about 12\% (15\%) at $p_{\rm T}$ = 3 GeV/$c$ to 5\% (7\%) 
at $p_{\rm T}$ = 10 GeV/$c$ in the 0--10\% (20--40\%) centrality class. 

The systematic uncertainties 
affecting the 
estimated $v_2^{\mu \leftarrow \pi, {\rm K}}$ are summarized in 
Table~2. 
They originate from i) the method used to 
measure the charged-particle $v_2^{\rm ch}$ 
in ATLAS, ii) the $\eta$ and $p_{\rm T}$ extrapolation of $v_2^{\rm ch}$ 
and iii) the treatment of the charged-particle
$v_2^{\rm ch}$ in the fast 
simulation procedure. 
As the event plane method was used for the $v_2^{\rm ch}$ measurement 
in ATLAS, the results range between the mean 
($\langle v_2^{\rm ch} \rangle$) and R.M.S.
($\sqrt {\langle (v_2^{\rm ch})^2 \rangle}$) 
of the true $v_2^{\rm ch}$ values due to fluctuations,
depending on the event plane resolution which varies with the 
collision centrality~\cite{Luzum:2012da}. 
According to a Monte-Carlo Glauber model~\cite{Luzum:2012da}, the ratio 
$\sqrt {\langle v_2^2 \rangle}/\langle v_2 \rangle$ is expected to vary 
from about 1.06 to 1.15. Consequently, a conservative systematic uncertainty 
of 15\% is applied to account for this bias and is propagated to 
$v_2^{\mu \leftarrow {\rm \pi, K}}$. 
The systematic uncertainty 
due to the $\eta$ 
extrapolation of $v_2^{\rm ch}$ is evaluated 
using several fit functions (first and 
third order polynomials, and Gaussian function) in the 
region 
$p_{\rm T} < 7$~GeV/$c$, and for larger $p_{\rm T}$ values 
an additional systematic uncertainty due to the extrapolation 
procedure is considered. The latter is determined by 
comparing the results 
obtained with the two extrapolation functions used in the interval 
$p_{\rm T} > 7$~GeV/$c$.
The systematic uncertainty due to the assumption 
on $v_2^{\rm ch}$ in the region $p_{\rm T} > 10$~GeV/$c$ is 
estimated by varying $v_2^{\rm ch}$ between 0 and the value in 
$10 < p_{\rm T} < 12$~GeV/$c$ in the fast simulations. 
Such uncertainty affects mainly the high $p_{\rm T}$ region 
($p_{\rm T} > 7$~GeV/$c$). 
Finally, the systematic uncertainty obtained by treating 
charged particles separately as pions and kaons is found to be negligible.
The various systematic uncertainty sources are propagated 
to the estimated $v_2^{\mu \leftarrow {\rm \pi, K}}$ and added in quadrature.

\begin{table}[hbt!]
\centering
\begin{tabular}{ c|c } \hline
Source & Systematic uncertainty (\%)\\ \hline
Input $v_2^{\rm ch}$ bias  & 9 \\
$v_2^{\rm ch}$ $\eta$ extrapolation  & 9--12 \\
$v_2^{\rm ch}$ high $p_{\rm T}$ extrapolation & 13--15 \\
$\pi$ and $\rm K$ in fast simulations & $<$ 1 \\ \hline
\end{tabular}
\caption{Systematic uncertainty sources affecting the 
estimated $v_2^{\mu \leftarrow {\rm \pi, K}}$ for the interval 
$3 < p_{\rm T} < 10$~GeV/$c$. They are stated as a percentage 
of the $v_2$ value. The given range reflects the 
dependence on the collision centrality. }
\label{tab:systv2decay}
\end{table}

The systematic uncertainty on $f^{\mu \leftarrow {\rm \pi,K}}$, detailed 
in~\cite{Abelev:2012qh}, includes the uncertainty 
on the generated $p_{\rm T}$ 
distributions of muons from charged pion and kaon decays, and the uncertainty 
on the measured inclusive muon $p_{\rm T}$ distributions. The former
originates from the input charged pion and kaon distributions, the rapidity 
extrapolation and the absorber effect. 
The systematic uncertainty 
on the measured inclusive muon yields contains the systematic 
uncertainty on detector response, 
residual mis-alignment and centrality dependence 
of the efficiency. 
This gives a total systematic uncertainty on 
$f^{\mu \leftarrow {\rm \pi,K}}$ of about 21\%
in the interval $3 <p_{\rm T} < 4.5$~GeV/$c$ with 
almost no dependence on the collision centrality. 
Finally, as done for the measurement of the heavy-flavour decay muon 
$R_{\rm AA}$~\cite{Abelev:2012qh}, the systematic 
uncertainty due to the unknown 
suppression of charged particles at forward rapidity is 
calculated by varying $f^{\mu \leftarrow {\rm \pi,K}}$ from 0 to two times 
the estimated value. This corresponds to a variation of 
$R_{\rm AA}^{\pi,{\rm K}}(p_{\rm T})$ at forward rapidity 
from 0 up to two times 
$[R_{\rm AA}^{\pi,{\rm K}}(p_{\rm T})]_{y = 0}$.
This systematic uncertainty amounts to 10--30\% in the interval 
$3 <p_{\rm T} < 4.5$~GeV/$c$, depending on the collision centrality and 
the flow analysis method.

Finally, the systematic uncertainty on the elliptic flow of 
muons from heavy-flavour 
decays, $v_2^{\mu \leftarrow {\rm HF}}$, contains two contributions: the 
systematic uncertainties on $v_2^\mu$, $v_2^{\mu \leftarrow {\rm \pi,K}}$ and 
$f^{\mu \leftarrow {\rm \pi,K}}$ propagated 
according to the definition of $v_2^{\mu \leftarrow HF}$ given in 
Eq.~(\ref{eq:v2HFdef}), and the systematic 
uncertainty due to the unknown 
suppression of charged particles at forward rapidity. 
The final systematic uncertainty on 
$v_2^{\mu \leftarrow {\rm HF}}$ is obtained by adding in quadrature the 
two contributions. It amounts to about 12\%--36\% in the interval 
$3 <p_{\rm T} < 4.5$~GeV/$c$, depending on the collision centrality and the 
flow analysis method.

\section{Results}\label{sec:results}
\begin{figure}[t]
\centering
\includegraphics[width=.49\textwidth]{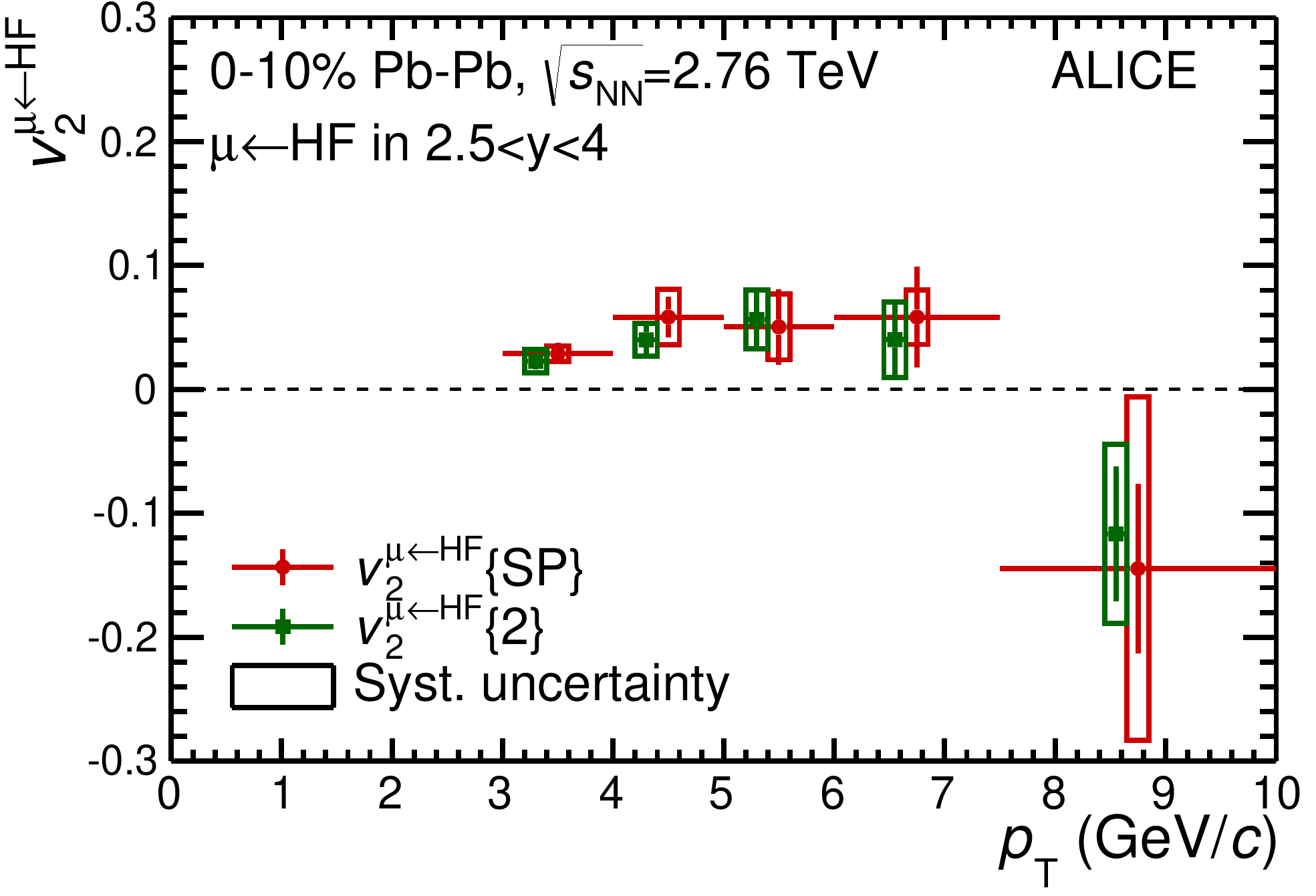}
\includegraphics[width=.49\textwidth]{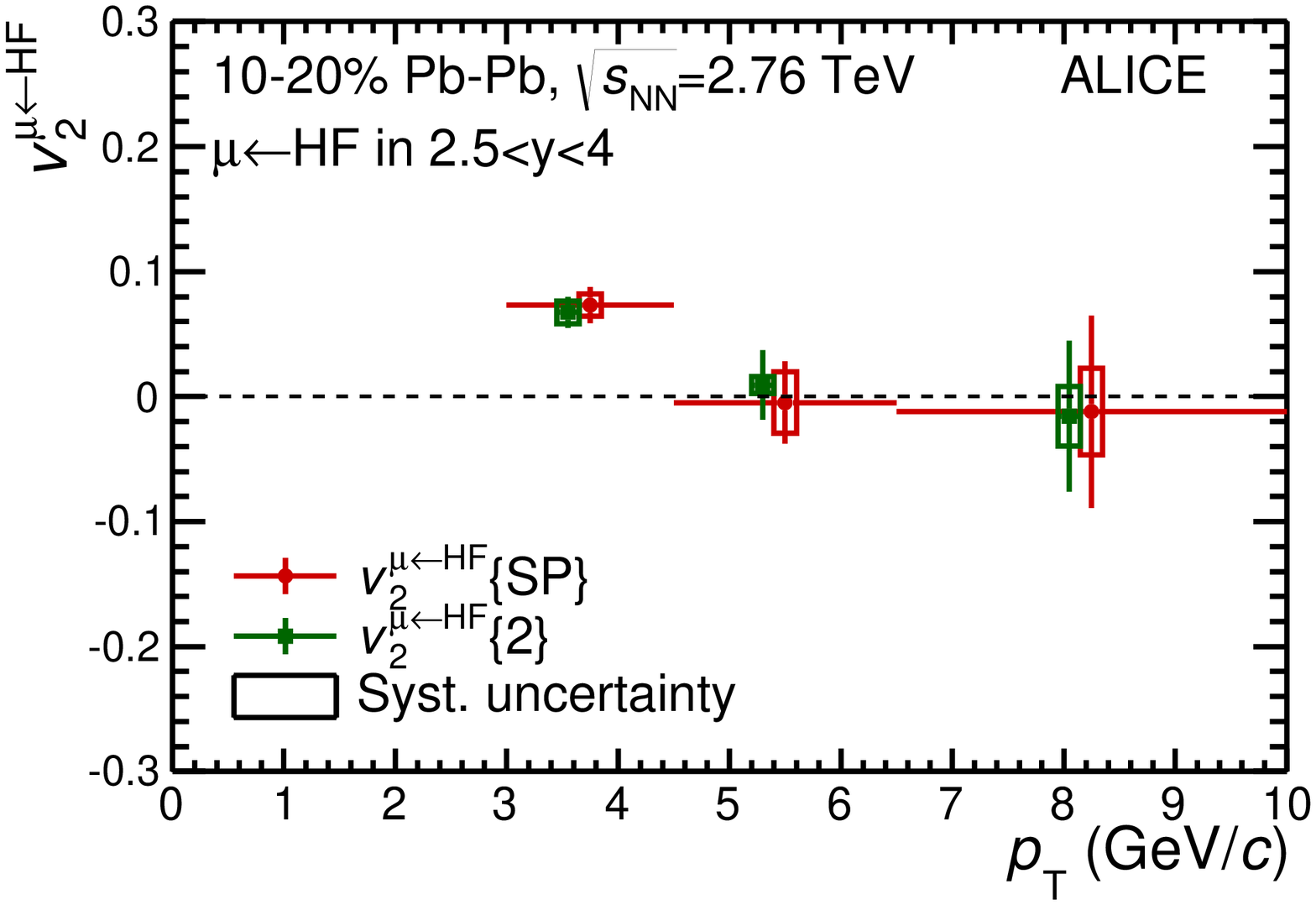}
\includegraphics[width=.49\textwidth]{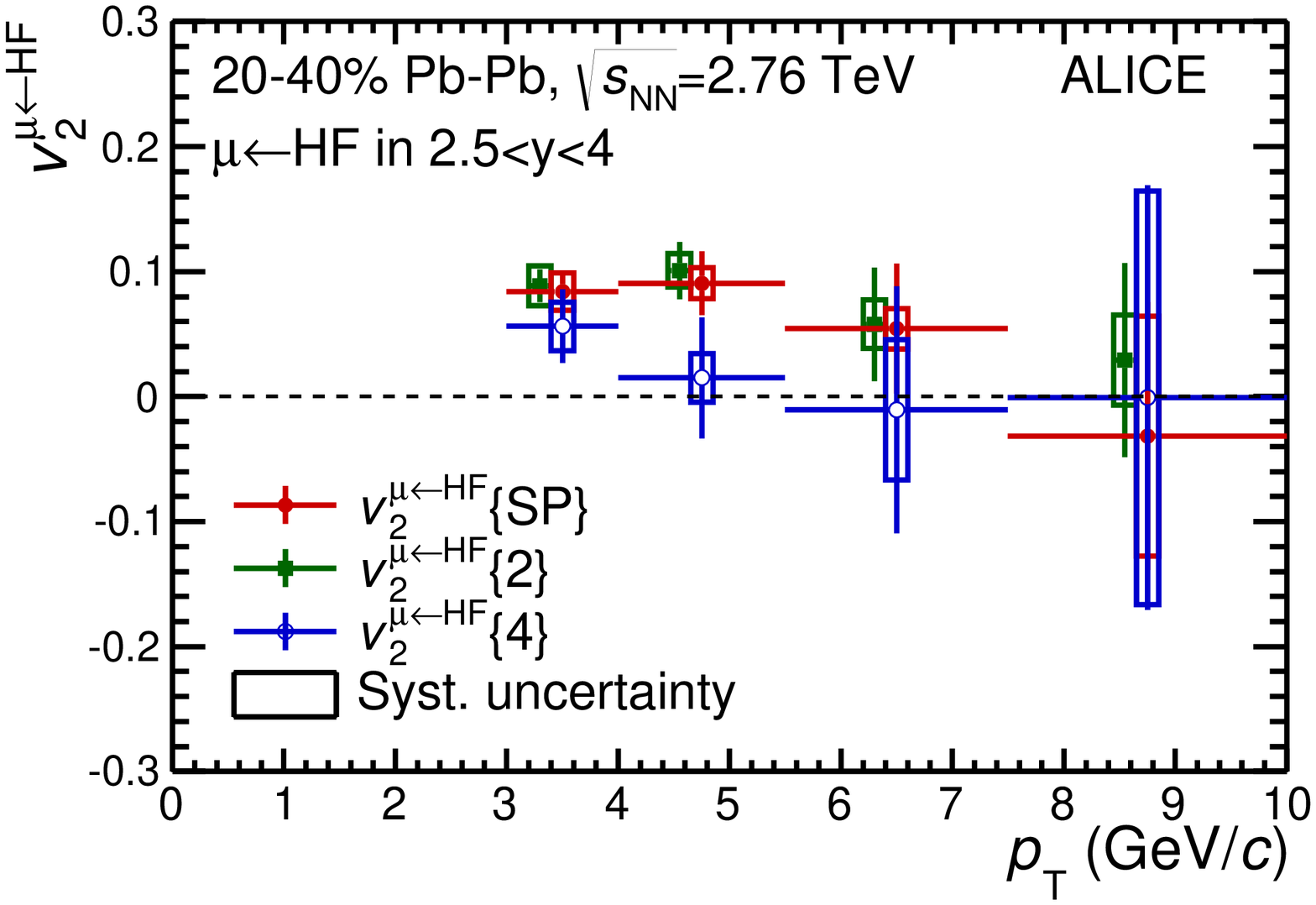}
\includegraphics[width=.49\textwidth]{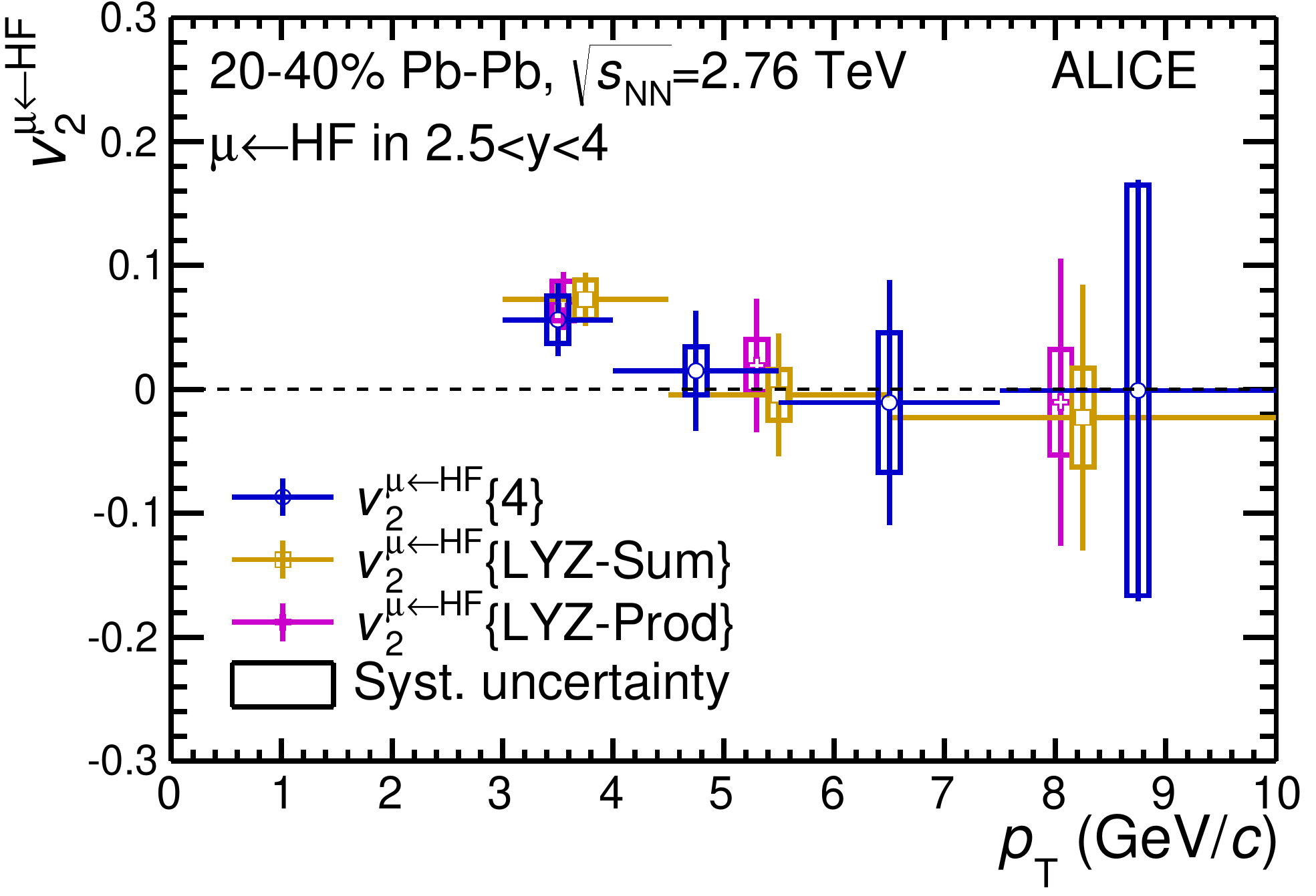}
\caption{
$p_{\rm T}$-differential elliptic flow of muons from heavy-flavour decays, 
$v_2^{\mu \leftarrow {\rm HF}}$, in $2.5 < y <4$ 
and various centrality intervals, in Pb--Pb collisions at 
$\sqrt {s_{\rm NN}}$ = 2.76 TeV. 
The symbols are placed at the centre of 
the $p_{\rm T}$ interval and, for visibility, 
the points from two-particle $Q$ 
cumulants and Lee-Yang zeros with product generating function 
are shifted horizontally. 
The meaning of the symbols is the same as in Fig.~\ref{fig:v2Mu}. 
The horizontal error bars are not plotted for shifted data points. 
The $p_{\rm T}$ intervals used with the Lee-Yang zeros 
method are different with respect to the other methods.
Upper panels: results from two-particle 
correlation flow methods (scalar product and two-particle $Q$ cumulants) 
in the 0--10\% (left) and 10--20\% (right) centrality intervals. 
Lower panels: results in the 20--40\% centrality interval 
from two-particle 
correlation flow methods (scalar product and two-particle $Q$ cumulants) and 
from four-particle $Q$ cumulants (left), and from 
four-particle $Q$ cumulants and Lee-Yang zeros (right). 
See the text for details.}
\label{fig:v2HFMu} 
\end{figure}

Figure~\ref{fig:v2HFMu} presents the $p_{\rm T}$-differential elliptic flow of 
muons from heavy-flavour hadron decays, 
$v_2^{\mu \leftarrow {\rm HF}}$, calculated 
with Eq.~(\ref{eq:v2HFdef}). The results are 
shown for the 
0--10\% (upper, left), 10--20\% (upper, right) and 20--40\% (bottom) 
centrality classes using the 
same flow methods as for the measurement of the inclusive muon elliptic flow 
(Fig.~\ref{fig:v2Mu}). 
When comparing the results to those 
obtained for inclusive muons (Fig.~\ref{fig:v2Mu}), one can notice that 
$v_2^{\mu \leftarrow {\rm HF}}$ and $v_2^\mu$ are similar
due to the small background fraction
(5\% to 15\%) in the $p_{\rm T}$ interval 3--10~GeV/$c$. 
The differences between the various methods 
are similar to those discussed for the measurement of the 
inclusive 
muon $v_2^\mu$ i.e. i) scalar product and two-particle $Q$ cumulants give 
compatible results, ii) consistent results are also 
found with four-particle $Q$ cumulants and Lee-Yang zeros, and iii) the 
$v_2^{\mu \leftarrow {\rm HF}}$ values extracted from 
these multi-particle correlation methods are smaller, although still 
compatible within uncertainties, than the ones 
obtained with two-particle correlation methods. As mentioned in 
Section~\ref{sec:inclv2}, such differences are 
expected if initial-state fluctuations play a role in the development of 
the final momentum-space anisotropy.

A positive $v_2^{\mu \leftarrow {\rm HF}}$ is observed at intermediate 
$p_{\rm T}$ for the 20--40\% and 10--20\% centrality classes
with a significance 
larger than $3\sigma$ when combining statistical and 
systematic uncertainties. 
In the 20--40\% centrality class, 
the values of the significance in the 
interval $3 < p_{\rm T} < 4$~GeV/$c$ ($4 < p_{\rm T} < 5.5$~GeV/$c$) 
are 4~$\sigma$ ( 3.2~$\sigma$) and 4.3~$\sigma$ (3.8~$\sigma$) 
with scalar product and two-particle $Q$ cumulants, respectively.
In the 10--20\% centrality class and in the interval 
$3 < p_{\rm T} < 4.5$~GeV/$c$, the values of the significance correspond to 
4.4$\sigma$ both with scalar product and two-particle $Q$ cumulants.
This behaviour results from the interplay between the significant interaction 
of heavy quarks with the expanding medium and the path-length 
dependence of in-medium 
parton energy loss~\cite{Gyulassy:2000gk,Shuryak:2001me}. 
The $v_2^{\mu \leftarrow {\rm HF}}$ of muons from heavy-flavour 
hadron decays decreases with increasing $p_{\rm T}$ and
becomes compatible with zero in the high $p_{\rm T}$ region. 

\begin{figure}[!t]
\centering
\includegraphics[width=.8\textwidth]{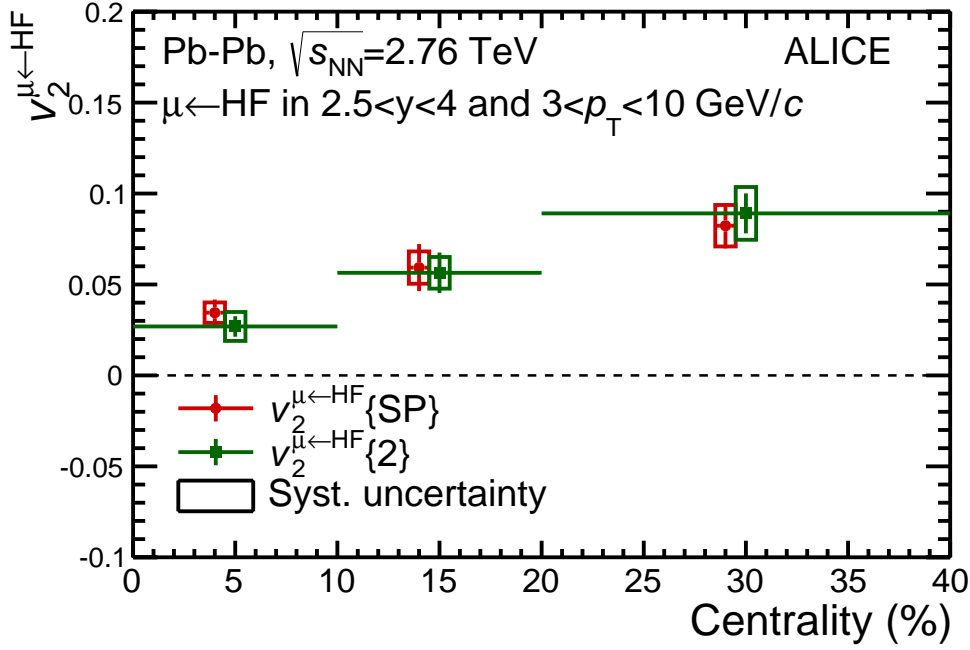}
\caption{Elliptic flow
of muons from heavy-flavour 
hadron decays as a function of the 
collision centrality in $2.5 < y <4$ and $3 < p_{\rm T} < 10$~GeV/$c$, 
in Pb--Pb collisions at $\sqrt {s_{\rm NN}}$ = 2.76 TeV. The results are 
obtained with scalar product and two-particle $Q$ cumulants. 
Vertical bars (open boxes) represent the statistical 
(systematic) uncertainty, the horizontal error bars correspond 
to the width of the centrality bin.
For visibility, the points from scalar product 
are shifted horizontally and the horizontal error bars 
are not plotted}.
\label{fig:v2HFMuIntCentQC2} 
\end{figure}

Figure~\ref{fig:v2HFMuIntCentQC2} shows the centrality 
dependence 
of the $p_{\rm T}$-integrated ($3 < p_{\rm T} < 10$~GeV/$c$) 
elliptic flow of muons 
from heavy-flavour hadron decays. It is investigated with 
scalar product and two-particle $Q$ cumulants, which can be 
applied in a wider event-multiplicity (i.e. centrality) interval compared 
to multi-particle correlation methods.
A significant decrease of the $v_2$ magnitude towards central collisions is 
observed. This 
is expected from the decrease of the initial spatial anisotropy 
from semi-central to central collisions.

ALICE has measured the elliptic flow 
of prompt D mesons 
in $\vert y \vert < 0.8$ in three centrality classes in 
the interval 0--50\% with various two-particle 
correlation methods~\cite{Abelev:2013lca,Abelev:2014ipa}. 
Similar trends as those reported here 
for muons from heavy-flavour decays are observed, although in different 
$p_{\rm T}$ and rapidity intervals. In particular, 
a positive $v_2$ was
observed for D mesons
in semi-central collisions in 2~$<$~$p_{\rm T}$~$<$~6~GeV/$c$.

The positive elliptic flow of muons from heavy-flavour 
hadron decays has been observed 
in a $p_{\rm T}$ interval from 3 to about 5~GeV/$c$
where the charm contribution is expected to be dominant with respect 
to the beauty component according to perturbative QCD 
calculations~\cite{Abelev:2012qh}. 
This 
measurement supports the interpretation of the ${\rm J}/\psi$ positive $v_2$ at 
forward rapidity~\cite{ALICE:2013xna} 
in terms of a significant contribution to
$\rm J/\psi$ production 
from recombination of flowing charm quarks in the deconfined medium.

\section{Comparison with models}\label{sec:models}

The results presented in this publication 
may constrain models describing the interactions of heavy quarks 
with the medium via elastic (collisional) 
and inelastic (radiative) processes, 
and in particular the 
parton energy loss dependence on the path-length within the medium. 

The elliptic flow coefficient
and the nuclear modification factor of muons from 
heavy-flavour hadron 
decays~\cite{Abelev:2012qh} are compared to 
the following three models.
The MC@sHQ + EPOS transport model~\cite{Nahrgang:2013xaa} 
treats the propagation 
of heavy quarks in the medium including collisional and 
radiative energy loss, within a 3 + 1 dimensional fluid dynamical expansion 
based on the EPOS model~\cite{Werner:2010aa,Werner:2012xh}. The 
hadronization of heavy quarks takes place at the transition temperature via 
recombination at low $p_{\rm T}$ and 
fragmentation at intermediate and 
high $p_{\rm T}$. The final-state hadronic interactions are not included in 
the model. TAMU~\cite{He:2014cla} is a transport model including only 
collisional processes via the Langevin equation.
The hydrodynamical expansion is constrained by $p_{\rm T}$ spectra and 
elliptic flow data of light-flavour hadrons. The hadronization
is modeled including
a component of recombination of heavy quarks with light-flavour 
hadrons in the QGP. 
The diffusion of 
heavy-flavour mesons in the hadronic phase is also included. 
BAMPS~\cite{Uphoff:2011ad,Fochler:2011en,Uphoff:2012gb} is a partonic 
transport model 
based on the Boltzmann approach to multi-parton 
scatterings. It includes collisional processes with a running strong 
coupling constant. The lack of radiative contributions is accounted for by 
scaling the binary cross section with a correction factor, 
tuned to describe the nuclear modification factor and elliptic flow results 
at RHIC energies. Vacuum fragmentation functions are used for 
the hadronization. 

\begin{figure}[!t]
\centering
\includegraphics[width=.49\textwidth]{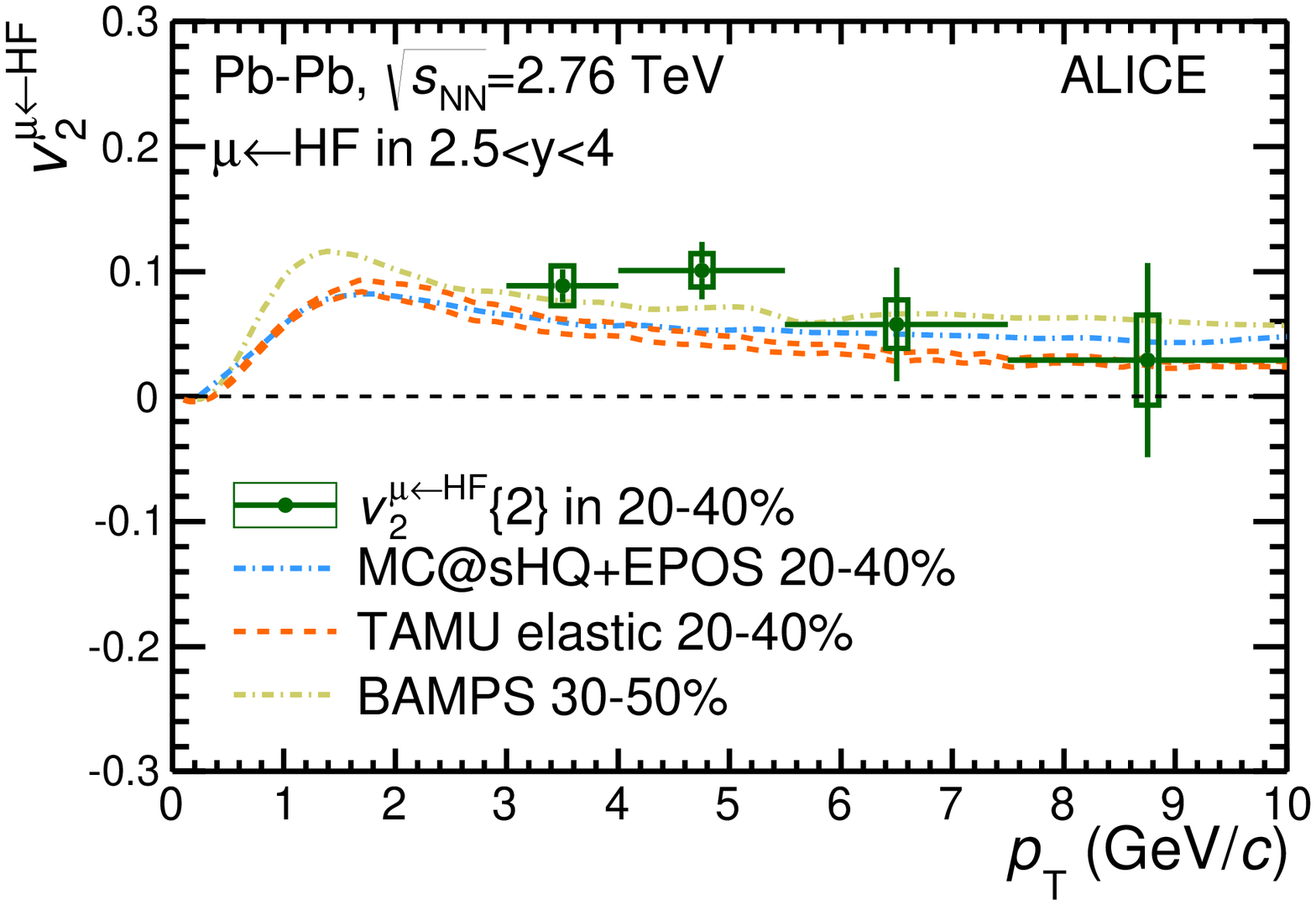}
\includegraphics[width=.49\textwidth]{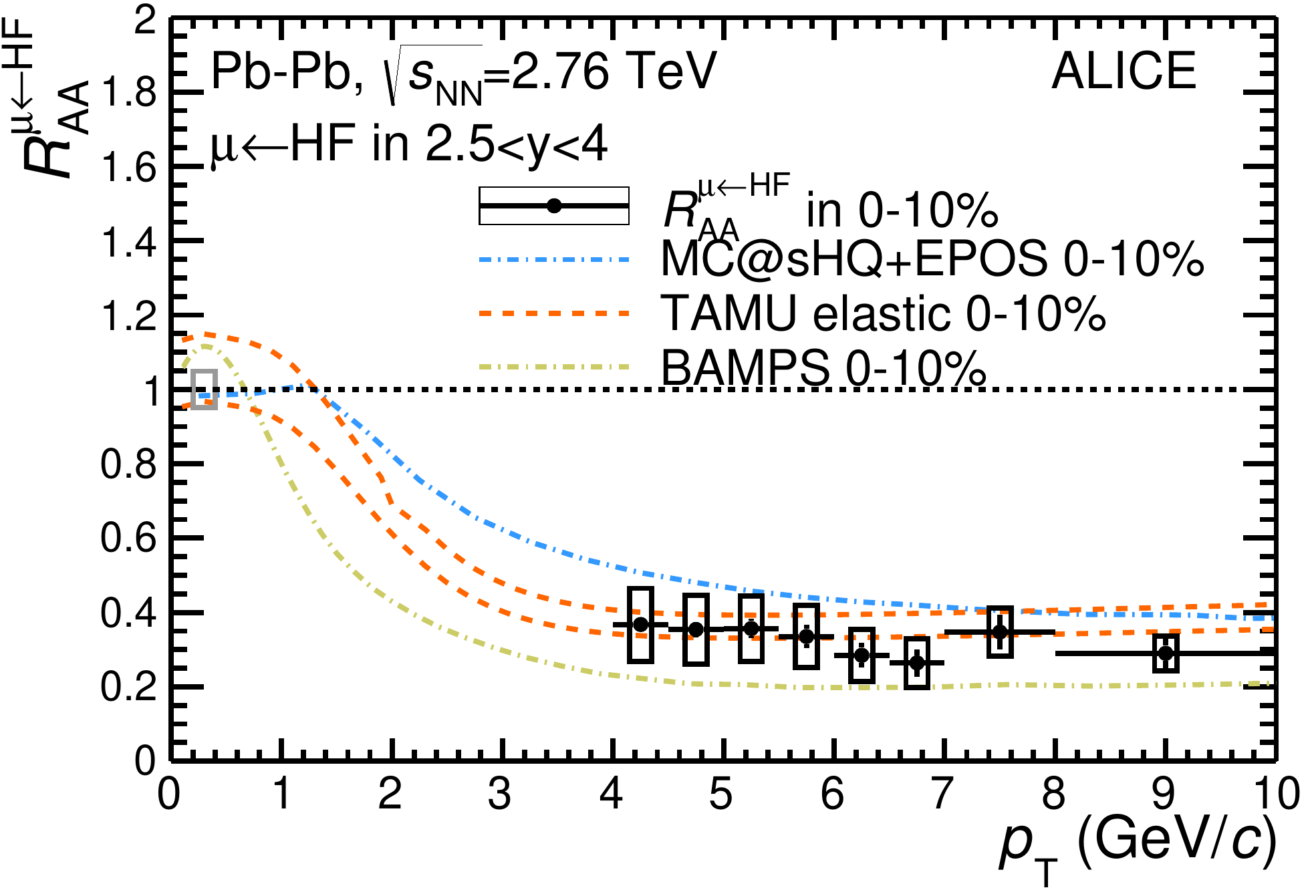}
\caption{Left: $p_{\rm T}$-differential 
elliptic flow of muons 
from heavy-flavour hadron decays 
in $2.5 < y <4$, in Pb--Pb collisions at $\sqrt {s_{\rm NN}}$ = 2.76 TeV 
for the centrality class 20--40\% compared to 
various transport model predictions:
MC@sHQ + EPOS~\cite{Nahrgang:2013xaa,Werner:2010aa,Werner:2012xh}, 
TAMU~\cite{He:2014cla} and 
BAMPS~\cite{Uphoff:2011ad,Fochler:2011en,Uphoff:2012gb}.
The TAMU model is shown with a theoretical uncertainty band.
Right: $p_{\rm T}$-differential $R_{\rm AA}$ of muons from heavy-flavour 
hadron decays for the centrality class 0--10\% from~\cite{Abelev:2012qh} compared 
to the same models as for $v_2^{\mu \leftarrow {\rm HF}}$.}
\label{fig:v2HFMuQC2Mod20-40}
\end{figure}

Figure~\ref{fig:v2HFMuQC2Mod20-40} (left) shows that the 
$p_{\rm T}$-differential elliptic flow
of muons from heavy-flavour hadron decays in the 
20--40\% centrality class is described reasonably well by the three 
models. However,
the BAMPS model tends to slightly underestimate the $R_{\rm AA}$ 
of muons from heavy-flavour decays in the 10\% most central collisions,
while the MC@sHQ$+$EPOS model tends to overestimate it.
The TAMU model
describes the $R_{\rm AA}$ measurement over the entire $p_{\rm T}$ interval 
within uncertainties and tends to slightly underestimates the 
$v_2^{\mu \leftarrow {\rm HF}}$ measurement in the low $p_{\rm T}$ region.
This indicates that it is challenging to simultaneously describe the 
strong suppression of high-$p_{\rm T}$
muons from heavy-flavour hadron decays in central collisions 
and the azimuthal anisotropy in semi-central collisions. 
Similar trends are also observed in the mid-rapidity 
region from the 
comparison of the $R_{\rm AA}$ and $v_2$ of D mesons with 
model calculations ~\cite{Abelev:2014ipa}. 

\section{Conclusions}\label{sec:conc}

In summary, we have reported on a measurement of 
the elliptic flow of 
muons from heavy-flavour hadron
decays at forward rapidity in central and 
semi-central Pb--Pb collisions at $\sqrt{s_{\rm NN}}$ = 2.76 TeV 
with the ALICE detector at the LHC. 

Measurements have been carried out using several methods which 
exhibit different
sensitivity to initial-state fluctuations and non-flow correlations.
The systematic comparison of scalar product, 
two- and four-particle $Q$ cumulants and Lee-Yang zeros helps 
in understanding the processes that build up the observed differences between 
two-particle correlation methods and multi-particle correlation methods 
and suggests that flow fluctuations are significant. 

The magnitude of 
the elliptic flow of muons from heavy-flavour hadron decays
increases from central to semi-central collisions and decreases with 
increasing $p_{\rm T}$, becoming compatible with zero at high $p_{\rm T}$
The results indicate 
a positive elliptic flow
with the scalar product and two-particle 
$Q$ cumulants in semi-central collisions (10--20\% and 20--40\% centrality 
classes) for the $p_{\rm T}$ interval from 3 to about 5 GeV/$c$
with a significance larger than $3\sigma$
The data are described by transport 
model calculations within uncertainties, although a 
simultaneous description of $R_{\rm AA}$
and $v_{2}$ remains a challenge.
The results reported in this Letter in 
various centrality classes may provide further 
important constraints to the models.


\newenvironment{acknowledgement}{\relax}{\relax}
\begin{acknowledgement}
\section*{Acknowledgements}
The ALICE Collaboration would like to thank all its engineers and technicians for their invaluable contributions to the construction of the experiment and the CERN accelerator teams for the outstanding performance of the LHC complex.
The ALICE Collaboration gratefully acknowledges the resources and support provided by all Grid centres and the Worldwide LHC Computing Grid (WLCG) collaboration.
The ALICE Collaboration acknowledges the following funding agencies for their support in building and
running the ALICE detector:
State Committee of Science,  World Federation of Scientists (WFS)
and Swiss Fonds Kidagan, Armenia,
Conselho Nacional de Desenvolvimento Cient\'{\i}fico e Tecnol\'{o}gico (CNPq), Financiadora de Estudos e Projetos (FINEP),
Funda\c{c}\~{a}o de Amparo \`{a} Pesquisa do Estado de S\~{a}o Paulo (FAPESP);
National Natural Science Foundation of China (NSFC), the Chinese Ministry of Education (CMOE)
and the Ministry of Science and Technology of China (MSTC);
Ministry of Education and Youth of the Czech Republic;
Danish Natural Science Research Council, the Carlsberg Foundation and the Danish National Research Foundation;
The European Research Council under the European Community's Seventh Framework Programme;
Helsinki Institute of Physics and the Academy of Finland;
French CNRS-IN2P3, the `Region Pays de Loire', `Region Alsace', `Region Auvergne' and CEA, France;
German Bundesministerium fur Bildung, Wissenschaft, Forschung und Technologie (BMBF) and the Helmholtz Association;
General Secretariat for Research and Technology, Ministry of
Development, Greece;
Hungarian Orszagos Tudomanyos Kutatasi Alappgrammok (OTKA) and National Office for Research and Technology (NKTH);
Department of Atomic Energy and Department of Science and Technology of the Government of India;
Istituto Nazionale di Fisica Nucleare (INFN) and Centro Fermi -
Museo Storico della Fisica e Centro Studi e Ricerche "Enrico
Fermi", Italy;
MEXT Grant-in-Aid for Specially Promoted Research, Ja\-pan;
Joint Institute for Nuclear Research, Dubna;
National Research Foundation of Korea (NRF);
Consejo Nacional de Cienca y Tecnologia (CONACYT), Direccion General de Asuntos del Personal Academico(DGAPA), M\'{e}xico, :Amerique Latine Formation academique â€“ European Commission(ALFA-EC) and the EPLANET Program
(European Particle Physics Latin American Network)
Stichting voor Fundamenteel Onderzoek der Materie (FOM) and the Nederlandse Organisatie voor Wetenschappelijk Onderzoek (NWO), Netherlands;
Research Council of Norway (NFR);
National Science Centre, Poland;
Ministry of National Education/Institute for Atomic Physics and Consiliul NaÅ£ional al CercetÄƒrii ÅžtiinÅ£ifice - Executive Agency for Higher Education Research Development and Innovation Funding (CNCS-UEFISCDI) - Romania;
Ministry of Education and Science of Russian Federation, Russian
Academy of Sciences, Russian Federal Agency of Atomic Energy,
Russian Federal Agency for Science and Innovations and The Russian
Foundation for Basic Research;
Ministry of Education of Slovakia;
Department of Science and Technology, South Africa;
Centro de Investigaciones Energeticas, Medioambientales y Tecnologicas (CIEMAT), E-Infrastructure shared between Europe and Latin America (EELA), Ministerio de Econom\'{i}a y Competitividad (MINECO) of Spain, Xunta de Galicia (Conseller\'{\i}a de Educaci\'{o}n),
Centro de Aplicaciones TecnolÃ³gicas y Desarrollo Nuclear (CEA\-DEN), Cubaenerg\'{\i}a, Cuba, and IAEA (International Atomic Energy Agency);
Swedish Research Council (VR) and Knut $\&$ Alice Wallenberg
Foundation (KAW);
Ukraine Ministry of Education and Science;
United Kingdom Science and Technology Facilities Council (STFC);
The United States Department of Energy, the United States National
Science Foundation, the State of Texas, and the State of Ohio;
Ministry of Science, Education and Sports of Croatia and  Unity through Knowledge Fund, Croatia.
Council of Scientific and Industrial Research (CSIR), New Delhi, India

\end{acknowledgement}

\bibliographystyle{utphys.bst}
\bibliography{AliFlowHFM}

\newpage
\appendix
\section{The ALICE Collaboration}
\label{app:collab}



\begingroup
\small
\begin{flushleft}
J.~Adam\Irefn{org40}\And
D.~Adamov\'{a}\Irefn{org83}\And
M.M.~Aggarwal\Irefn{org87}\And
G.~Aglieri Rinella\Irefn{org36}\And
M.~Agnello\Irefn{org110}\And
N.~Agrawal\Irefn{org48}\And
Z.~Ahammed\Irefn{org131}\And
S.U.~Ahn\Irefn{org68}\And
S.~Aiola\Irefn{org135}\And
A.~Akindinov\Irefn{org58}
% ...
\renewcommand\labelenumi{\textsuperscript{\theenumi}~}

\section*{Affiliation notes}
\renewcommand\theenumi{\roman{enumi}}
\begin{Authlist}
\item \Adef{0}Deceased
\item \Adef{idp3777184}{Also at: M.V. Lomonosov Moscow State University, D.V. Skobeltsyn Institute of Nuclear, Physics, Moscow, Russia}
\item \Adef{idp5932384}{Also at: University of Kansas, Lawrence, Kansas, United States}
\end{Authlist}

\section*{Collaboration Institutes}
\renewcommand\theenumi{\arabic{enumi}~}
\begin{Authlist}

\item \Idef{org1}A.I. Alikhanyan National Science Laboratory (Yerevan Physics Institute) Foundation, Yerevan, Armenia
\item \Idef{org2}Benem\'{e}rita Universidad Aut\'{o}noma de Puebla, Puebla, Mexico
\item \Idef{org3}Bogolyubov Institute for Theoretical Physics, Kiev, Ukraine
\item \Idef{org4}Bose Institute, Department of Physics and Centre for Astroparticle Physics and Space Science (CAPSS), Kolkata, India
\item \Idef{org5}Budker Institute for Nuclear Physics, Novosibirsk, Russia
\item \Idef{org6}California Polytechnic State University, San Luis Obispo, California, United States
\item \Idef{org7}Central China Normal University, Wuhan, China
\item \Idef{org8}Centre de Calcul de l'IN2P3, Villeurbanne, France
\item \Idef{org9}Centro de Aplicaciones Tecnol\'{o}gicas y Desarrollo Nuclear (CEADEN), Havana, Cuba
\item \Idef{org10}Centro de Investigaciones Energ\'{e}ticas Medioambientales y Tecnol\'{o}gicas (CIEMAT), Madrid, Spain
\item \Idef{org11}Centro de Investigaci\'{o}n y de Estudios Avanzados (CINVESTAV), Mexico City and M\'{e}rida, Mexico
\item \Idef{org12}Centro Fermi - Museo Storico della Fisica e Centro Studi e Ricerche ``Enrico Fermi'', Rome, Italy
\item \Idef{org13}Chicago State University, Chicago, Illinois, USA
\item \Idef{org14}China Institute of Atomic Energy, Beijing, China
\item \Idef{org15}Commissariat \`{a} l'Energie Atomique, IRFU, Saclay, France
\item \Idef{org16}COMSATS Institute of Information Technology (CIIT), Islamabad, Pakistan
\item \Idef{org17}Departamento de F\'{\i}sica de Part\'{\i}culas and IGFAE, Universidad de Santiago de Compostela, Santiago de Compostela, Spain
\item \Idef{org18}Department of Physics and Technology, University of Bergen, Bergen, Norway
\item \Idef{org19}Department of Physics, Aligarh Muslim University, Aligarh, India
\item \Idef{org20}Department of Physics, Ohio State University, Columbus, Ohio, United States
\item \Idef{org21}Department of Physics, Sejong University, Seoul, South Korea
\item \Idef{org22}Department of Physics, University of Oslo, Oslo, Norway
\item \Idef{org23}Dipartimento di Elettrotecnica ed Elettronica del Politecnico, Bari, Italy
\item \Idef{org24}Dipartimento di Fisica dell'Universit\`{a} 'La Sapienza' and Sezione INFN Rome, Italy
\item \Idef{org25}Dipartimento di Fisica dell'Universit\`{a} and Sezione INFN, Cagliari, Italy
\item \Idef{org26}Dipartimento di Fisica dell'Universit\`{a} and Sezione INFN, Trieste, Italy
\item \Idef{org27}Dipartimento di Fisica dell'Universit\`{a} and Sezione INFN, Turin, Italy
\item \Idef{org28}Dipartimento di Fisica e Astronomia dell'Universit\`{a} and Sezione INFN, Bologna, Italy
\item \Idef{org29}Dipartimento di Fisica e Astronomia dell'Universit\`{a} and Sezione INFN, Catania, Italy
\item \Idef{org30}Dipartimento di Fisica e Astronomia dell'Universit\`{a} and Sezione INFN, Padova, Italy
\item \Idef{org31}Dipartimento di Fisica `E.R.~Caianiello' dell'Universit\`{a} and Gruppo Collegato INFN, Salerno, Italy
\item \Idef{org32}Dipartimento di Scienze e Innovazione Tecnologica dell'Universit\`{a} del  Piemonte Orientale and Gruppo Collegato INFN, Alessandria, Italy
\item \Idef{org33}Dipartimento Interateneo di Fisica `M.~Merlin' and Sezione INFN, Bari, Italy
\item \Idef{org34}Division of Experimental High Energy Physics, University of Lund, Lund, Sweden
\item \Idef{org35}Eberhard Karls Universit\"{a}t T\"{u}bingen, T\"{u}bingen, Germany
\item \Idef{org36}European Organization for Nuclear Research (CERN), Geneva, Switzerland
\item \Idef{org37}Excellence Cluster Universe, Technische Universit\"{a}t M\"{u}nchen, Munich, Germany
\item \Idef{org38}Faculty of Engineering, Bergen University College, Bergen, Norway
\item \Idef{org39}Faculty of Mathematics, Physics and Informatics, Comenius University, Bratislava, Slovakia
\item \Idef{org40}Faculty of Nuclear Sciences and Physical Engineering, Czech Technical University in Prague, Prague, Czech Republic
\item \Idef{org41}Faculty of Science, P.J.~\v{S}af\'{a}rik University, Ko\v{s}ice, Slovakia
\item \Idef{org42}Faculty of Technology, Buskerud and Vestfold University College, Vestfold, Norway
\item \Idef{org43}Frankfurt Institute for Advanced Studies, Johann Wolfgang Goethe-Universit\"{a}t Frankfurt, Frankfurt, Germany
\item \Idef{org44}Gangneung-Wonju National University, Gangneung, South Korea
\item \Idef{org45}Gauhati University, Department of Physics, Guwahati, India
\item \Idef{org46}Helsinki Institute of Physics (HIP), Helsinki, Finland
\item \Idef{org47}Hiroshima University, Hiroshima, Japan
\item \Idef{org48}Indian Institute of Technology Bombay (IIT), Mumbai, India
\item \Idef{org49}Indian Institute of Technology Indore, Indore (IITI), India
\item \Idef{org50}Inha University, Incheon, South Korea
\item \Idef{org51}Institut de Physique Nucl\'eaire d'Orsay (IPNO), Universit\'e Paris-Sud, CNRS-IN2P3, Orsay, France
\item \Idef{org52}Institut f\"{u}r Informatik, Johann Wolfgang Goethe-Universit\"{a}t Frankfurt, Frankfurt, Germany
\item \Idef{org53}Institut f\"{u}r Kernphysik, Johann Wolfgang Goethe-Universit\"{a}t Frankfurt, Frankfurt, Germany
\item \Idef{org54}Institut f\"{u}r Kernphysik, Westf\"{a}lische Wilhelms-Universit\"{a}t M\"{u}nster, M\"{u}nster, Germany
\item \Idef{org55}Institut Pluridisciplinaire Hubert Curien (IPHC), Universit\'{e} de Strasbourg, CNRS-IN2P3, Strasbourg, France
\item \Idef{org56}Institute for Nuclear Research, Academy of Sciences, Moscow, Russia
\item \Idef{org57}Institute for Subatomic Physics of Utrecht University, Utrecht, Netherlands
\item \Idef{org58}Institute for Theoretical and Experimental Physics, Moscow, Russia
\item \Idef{org59}Institute of Experimental Physics, Slovak Academy of Sciences, Ko\v{s}ice, Slovakia
\item \Idef{org60}Institute of Physics, Academy of Sciences of the Czech Republic, Prague, Czech Republic
\item \Idef{org61}Institute of Physics, Bhubaneswar, India
\item \Idef{org62}Institute of Space Science (ISS), Bucharest, Romania
\item \Idef{org63}Instituto de Ciencias Nucleares, Universidad Nacional Aut\'{o}noma de M\'{e}xico, Mexico City, Mexico
\item \Idef{org64}Instituto de F\'{\i}sica, Universidad Nacional Aut\'{o}noma de M\'{e}xico, Mexico City, Mexico
\item \Idef{org65}iThemba LABS, National Research Foundation, Somerset West, South Africa
\item \Idef{org66}Joint Institute for Nuclear Research (JINR), Dubna, Russia
\item \Idef{org67}Konkuk University, Seoul, South Korea
\item \Idef{org68}Korea Institute of Science and Technology Information, Daejeon, South Korea
\item \Idef{org69}KTO Karatay University, Konya, Turkey
\item \Idef{org70}Laboratoire de Physique Corpusculaire (LPC), Clermont Universit\'{e}, Universit\'{e} Blaise Pascal, CNRS--IN2P3, Clermont-Ferrand, France
\item \Idef{org71}Laboratoire de Physique Subatomique et de Cosmologie, Universit\'{e} Grenoble-Alpes, CNRS-IN2P3, Grenoble, France
\item \Idef{org72}Laboratori Nazionali di Frascati, INFN, Frascati, Italy
\item \Idef{org73}Laboratori Nazionali di Legnaro, INFN, Legnaro, Italy
\item \Idef{org74}Lawrence Berkeley National Laboratory, Berkeley, California, United States
\item \Idef{org75}Lawrence Livermore National Laboratory, Livermore, California, United States
\item \Idef{org76}Moscow Engineering Physics Institute, Moscow, Russia
\item \Idef{org77}National Centre for Nuclear Studies, Warsaw, Poland
\item \Idef{org78}National Institute for Physics and Nuclear Engineering, Bucharest, Romania
\item \Idef{org79}National Institute of Science Education and Research, Bhubaneswar, India
\item \Idef{org80}Niels Bohr Institute, University of Copenhagen, Copenhagen, Denmark
\item \Idef{org81}Nikhef, Nationaal instituut voor subatomaire fysica, Amsterdam, Netherlands
\item \Idef{org82}Nuclear Physics Group, STFC Daresbury Laboratory, Daresbury, United Kingdom
\item \Idef{org83}Nuclear Physics Institute, Academy of Sciences of the Czech Republic, \v{R}e\v{z} u Prahy, Czech Republic
\item \Idef{org84}Oak Ridge National Laboratory, Oak Ridge, Tennessee, United States
\item \Idef{org85}Petersburg Nuclear Physics Institute, Gatchina, Russia
\item \Idef{org86}Physics Department, Creighton University, Omaha, Nebraska, United States
\item \Idef{org87}Physics Department, Panjab University, Chandigarh, India
\item \Idef{org88}Physics Department, University of Athens, Athens, Greece
\item \Idef{org89}Physics Department, University of Cape Town, Cape Town, South Africa
\item \Idef{org90}Physics Department, University of Jammu, Jammu, India
\item \Idef{org91}Physics Department, University of Rajasthan, Jaipur, India
\item \Idef{org92}Physik Department, Technische Universit\"{a}t M\"{u}nchen, Munich, Germany
\item \Idef{org93}Physikalisches Institut, Ruprecht-Karls-Universit\"{a}t Heidelberg, Heidelberg, Germany
\item \Idef{org94}Purdue University, West Lafayette, Indiana, United States
\item \Idef{org95}Pusan National University, Pusan, South Korea
\item \Idef{org96}Research Division and ExtreMe Matter Institute EMMI, GSI Helmholtzzentrum f\"ur Schwerionenforschung, Darmstadt, Germany
\item \Idef{org97}Rudjer Bo\v{s}kovi\'{c} Institute, Zagreb, Croatia
\item \Idef{org98}Russian Federal Nuclear Center (VNIIEF), Sarov, Russia
\item \Idef{org99}Russian Research Centre Kurchatov Institute, Moscow, Russia
\item \Idef{org100}Saha Institute of Nuclear Physics, Kolkata, India
\item \Idef{org101}School of Physics and Astronomy, University of Birmingham, Birmingham, United Kingdom
\item \Idef{org102}Secci\'{o}n F\'{\i}sica, Departamento de Ciencias, Pontificia Universidad Cat\'{o}lica del Per\'{u}, Lima, Peru
\item \Idef{org103}Sezione INFN, Bari, Italy
\item \Idef{org104}Sezione INFN, Bologna, Italy
\item \Idef{org105}Sezione INFN, Cagliari, Italy
\item \Idef{org106}Sezione INFN, Catania, Italy
\item \Idef{org107}Sezione INFN, Padova, Italy
\item \Idef{org108}Sezione INFN, Rome, Italy
\item \Idef{org109}Sezione INFN, Trieste, Italy
\item \Idef{org110}Sezione INFN, Turin, Italy
\item \Idef{org111}SSC IHEP of NRC Kurchatov institute, Protvino, Russia
\item \Idef{org112}SUBATECH, Ecole des Mines de Nantes, Universit\'{e} de Nantes, CNRS-IN2P3, Nantes, France
\item \Idef{org113}Suranaree University of Technology, Nakhon Ratchasima, Thailand
\item \Idef{org114}Technical University of Ko\v{s}ice, Ko\v{s}ice, Slovakia
\item \Idef{org115}Technical University of Split FESB, Split, Croatia
\item \Idef{org116}The Henryk Niewodniczanski Institute of Nuclear Physics, Polish Academy of Sciences, Cracow, Poland
\item \Idef{org117}The University of Texas at Austin, Physics Department, Austin, Texas, USA
\item \Idef{org118}Universidad Aut\'{o}noma de Sinaloa, Culiac\'{a}n, Mexico
\item \Idef{org119}Universidade de S\~{a}o Paulo (USP), S\~{a}o Paulo, Brazil
\item \Idef{org120}Universidade Estadual de Campinas (UNICAMP), Campinas, Brazil
\item \Idef{org121}University of Houston, Houston, Texas, United States
\item \Idef{org122}University of Jyv\"{a}skyl\"{a}, Jyv\"{a}skyl\"{a}, Finland
\item \Idef{org123}University of Liverpool, Liverpool, United Kingdom
\item \Idef{org124}University of Tennessee, Knoxville, Tennessee, United States
\item \Idef{org125}University of the Witwatersrand, Johannesburg, South Africa
\item \Idef{org126}University of Tokyo, Tokyo, Japan
\item \Idef{org127}University of Tsukuba, Tsukuba, Japan
\item \Idef{org128}University of Zagreb, Zagreb, Croatia
\item \Idef{org129}Universit\'{e} de Lyon, Universit\'{e} Lyon 1, CNRS/IN2P3, IPN-Lyon, Villeurbanne, France
\item \Idef{org130}V.~Fock Institute for Physics, St. Petersburg State University, St. Petersburg, Russia
\item \Idef{org131}Variable Energy Cyclotron Centre, Kolkata, India
\item \Idef{org132}Warsaw University of Technology, Warsaw, Poland
\item \Idef{org133}Wayne State University, Detroit, Michigan, United States
\item \Idef{org134}Wigner Research Centre for Physics, Hungarian Academy of Sciences, Budapest, Hungary
\item \Idef{org135}Yale University, New Haven, Connecticut, United States
\item \Idef{org136}Yonsei University, Seoul, South Korea
\item \Idef{org137}Zentrum f\"{u}r Technologietransfer und Telekommunikation (ZTT), Fachhochschule Worms, Worms, Germany
\end{Authlist}
\endgroup


\end{document}